%% file: paper_arXiv.tex
\documentclass[
 reprint,
 nofootinbib,
 amsmath,amssymb,
 aps,
 pra,
 showkeys,
 longbibliography,
superscriptaddress,
]{revtex4-1}

\usepackage{graphicx} 
\usepackage{dcolumn} 
\usepackage{bm} 
\usepackage{amsmath}
\usepackage{siunitx}

\usepackage{gensymb}
\usepackage[symbol]{footmisc}
\usepackage{booktabs}
\usepackage{makecell}
\usepackage{threeparttable}
\usepackage{array}
\newcolumntype{P}[1]{>{\centering\arraybackslash}p{#1}}

\usepackage{geometry}
\usepackage{longtable}
\usepackage{booktabs}

\usepackage{makecell}

\begin{document}

\preprint{APS/123-QED}

\title{Nanophotonic thermal management in X-ray tubes}

\author{Simo Pajovic}
\affiliation{Department of Mechanical Engineering, Massachusetts Institute of Technology, Cambridge, MA, 02139, U.S.A.}
\email{pajovics@mit.edu}
\author{Charles Roques-Carmes}
\affiliation{Research Laboratory of Electronics, Massachusetts Institute of Technology, Cambridge, MA, 02139, U.S.A.}
\affiliation{E. L. Ginzton Laboratory, Stanford University, Stanford, CA, 94305, U.S.A.}
\author{Seou Choi}
\affiliation{Research Laboratory of Electronics, Massachusetts Institute of Technology, Cambridge, MA, 02139, U.S.A.}
\author{Steven E. Kooi}
\affiliation{Institute for Soldier Nanotechnologies, Massachusetts Institute of Technology, Cambridge, MA, 02139, U.S.A.}
\author{Rajiv~Gupta}
\affiliation{Department of Radiology, Massachusetts General Hospital, Boston, MA, 02114, U.S.A.}
\author{Michael E. Zalis}
\affiliation{Department of Radiology, Massachusetts General Hospital, Boston, MA, 02114, U.S.A.}
\author{Ivan Čelanović}
\affiliation{Institute for Soldier Nanotechnologies, Massachusetts Institute of Technology, Cambridge, MA, 02139, U.S.A.}
\author{Marin Soljačić}
\affiliation{Department of Physics, Massachusetts Institute of Technology, Cambridge, MA, 02139, U.S.A.}

\date{\today} 

\begin{abstract}
In X-ray tubes, more than 99\% of the kilowatts of power supplied to generate X-rays via bremsstrahlung are lost in the form of heat generation in the anode. Therefore, thermal management is a critical barrier to the development of more powerful X-ray tubes with higher brightness and spatial coherence, which are needed to translate imaging modalities such as phase-contrast imaging to the clinic. In rotating anode X-ray tubes, the most common design, thermal radiation is a bottleneck that prevents efficient cooling of the anode---the hottest part of the device by far. We predict that nanophotonically patterning the anode of an X-ray tube enhances heat dissipation via thermal radiation, enabling it to operate at higher powers without increasing in temperature. The focal spot size, which is related to the spatial coherence of generated X-rays, can also be made smaller at a constant temperature. A major advantage of our “nanophotonic thermal management” approach is that in principle, it allows for complete control over the spectrum and direction of thermal radiation, which can lead to optimal thermal routing and improved performance.
\end{abstract}

\keywords{X-Ray Tubes, Thermal Management, Nanophotonics, Thermal Radiation, X-Ray Imaging, High Temperature} 

\maketitle

X-ray imaging is one of the most important clinical tools for detection and diagnosis of disease: in the United States, over 80 million computed tomography (CT) scans are performed each year~\cite{Harvard2021}. Imaging modalities that require ultra-bright X-rays such as phase-contrast imaging (PCI) are of particular note because they can reveal detailed information about objects~\cite{Momose2005review,Bravin2012review}. However, many commercially available X-ray tubes lack the brightness and spatial coherence necessary to translate laboratory demonstrations to clinical applications. On the other hand, imaging modalities such as CT are demanding of X-ray tubes in almost every aspect: X-ray photon flux (brightness), photons per patient (dose), duration of scan, and even the photon energies used for imaging. At the same time, X-ray imaging consumes vast amounts of energy. A single-source CT scanner consumes about 25,000 kWh per year, with an additional 40\% for cooling the X-ray tube and computers required for image reconstruction---in total, the energy consumption of approximately three U.S. homes per year~\cite{GE2021}. Thus, although the deployment of these transformative technologies is attractive, it is also challenging due to demanding brightness and power requirements.

One way to address these challenges is thermal management. In an X-ray tube, electrons emitted by the cathode are accelerated across the vacuum of the housing by the tube current $I_t$ and voltage $V_t$ and collide with the anode, usually made of tungsten, to generate X-rays via bremsstrahlung. Less than 1\% of the power input, $I_{t}V_{t}$, on the order of tens of \unit{\kilo\watt}, is converted into X-rays; the rest is converted into heat. As a direct result of this, the majority of failure modes of X-ray tubes are thermal in origin, preventing high-power (i.e., high-brightness) operation. Importantly, thermal management also constrains the spatial resolution of X-ray imaging. The size of the focal spot from which X-rays emerge depends on the size of the electron beam, but as the focal spot size decreases, the local heat flux increases. This drives up the focal spot temperature, which is the hottest part of the anode and, therefore, the part most susceptible to failure modes such as melting and cracking~\cite{Iversen1984}. Because of the relatively low thermal conductivity of tungsten, traditional approaches to thermal management in X-ray tubes have focused on finding ways to redistribute heat, including by rotating the anode at high speeds (on the order of 10,000 rpm), liquid cooling, and combinations thereof. However, since the housing is evacuated, an important mechanism of heat dissipation is thermal radiation from the anode to the housing. Since the emissivity of tungsten is relatively low, the anode is a poor emitter, creating an obstacle to thermal management and achieving X-ray tubes with higher power and brightness. Despite this, engineering the emissivity of the anode remains an underexplored approach to thermal management in X-ray tubes.

In parallel to the development of thermal management in X-ray tubes, thermal management in systems that exchange heat via radiation has been revolutionized by photonic crystals designed to achieve a desired emissivity~\cite{Li2018}. The physics of thermal radiation control depend on the goal(s) of the design, such as spectral, angular, and/or polarization control. For example, when spectrally selective emissivity is desirable (e.g., thermophotovoltaics), the photonic crystal, typically consisting of a periodic array of cavities, satisfies Q-matching between absorption and confinement. Thus, thermal radiation below the cutoff wavelength can outcouple to free space, while that above it is resonantly absorbed~\cite{Yeng2012}. In this way, the critical difference between nanophotonics and X-ray tube designs that aim to maximize emissivity using bulk materials and coatings (such as graphite and ``black chrome''~\cite{Behling2021radiation}) is the ability to completely control the properties of thermal radiation. This has the potential to enable new X-ray tube designs that optimally route heat throughout the system, accounting for the particular spectral and directional characteristics of the materials used for building the X-ray tube.

Here, we propose the application of nanophotonics to tailor emissivity in X-ray tubes, where radiation is a dominant mode of heat transfer. We show that patterning a tungsten anode with a photonic crystal could lead to improvements in thermal management, which helps to alleviate both the demanding power requirements of X-ray tubes and the thermal bottleneck preventing higher doses and smaller focal spot sizes, which are important for certain imaging modalities. To illustrate the concept of ``nanophotonic thermal management'' in X-ray tubes, we use rigorous coupled wave analysis (RCWA) to calculate the spectral directional emissivity of a photonic crystal via reciprocity. Incorporating our nanophotonic design into a heat transfer model of an X-ray tube, we estimate the hottest (i.e., limiting) temperature in the X-ray tube as a function of the operating power. We predict that nanophotonic thermal management can lead to a $>$1.2$\times$ enhancement in the operating power of an X-ray tube at typical operating temperatures. We also show that our approach can be used to enable smaller focal spot sizes. Our approach makes use of nanophotonic designs that have been experimentally demonstrated in the context of high-temperature, thermochemically-stable thermal emitters for use with thermophotovoltaic cells~\cite{Celanovic2008,Yeng2012,Lenert2014,Chan2017,Sakakibara2022}. By taking advantage of an existing technology and directly patterning the anode (as opposed to using a coating), our approach to thermal management in X-ray tubes can lead to simpler anode designs that use fewer materials to achieve similar performance to high emissivity coatings~\cite{Behling2021radiation}. Overall, ``nanophotonic thermal management'' could help pave a way for the next generation of high-power, high-spatial-resolution sources in clinical applications.

\section*{Results and Discussion}

\begin{figure*}
  \includegraphics[width=\textwidth]{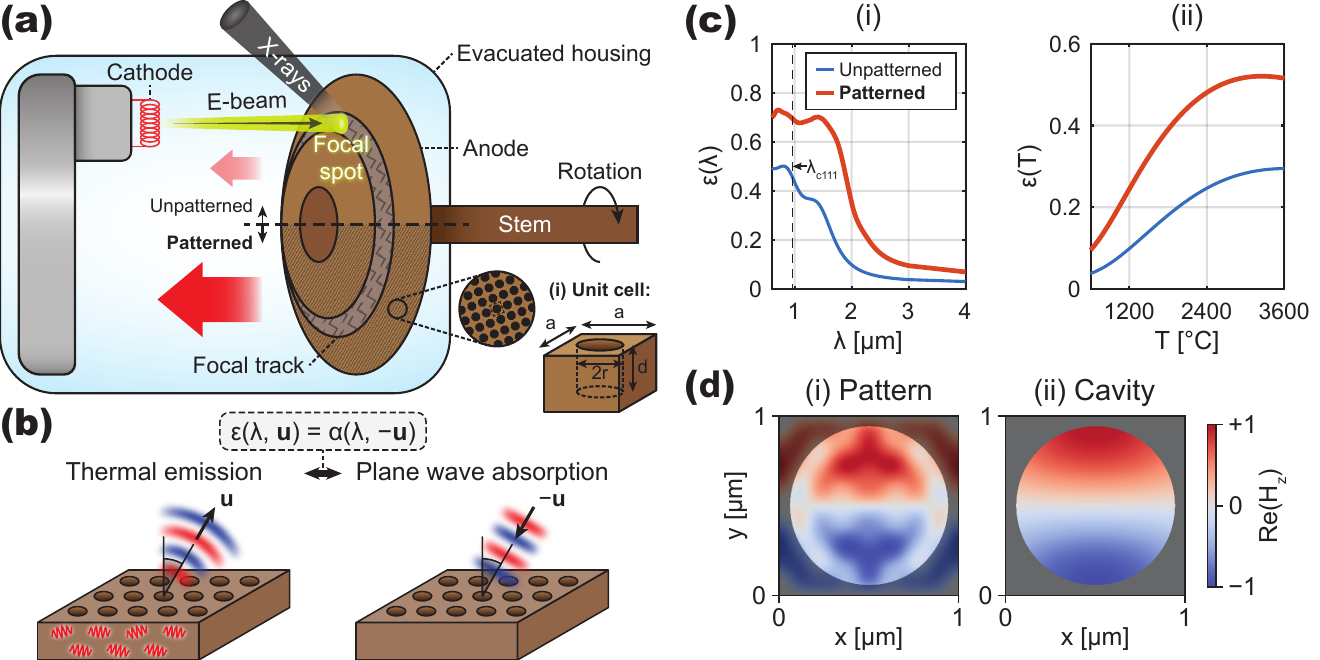}
  \caption{\textbf{Enhanced operating power in X-ray tubes via nanophotonic thermal management to tailor emissivity.} (a) Schematic of a rotating-anode X-ray tube with photonic crystal-patterned anode. The anode exchanges heat with its surroundings, namely the housing, via thermal radiation. By patterning the anode with a periodic array of holes commensurate with the peak wavelength of thermal radiation, radiative heat transfer is nanophotonically enhanced. (b) Kirchhoff's law of thermal radiation, or reciprocity: the spectral directional emissivity of the patterned surface equals its spectral directional absorptivity. (c) Emissivity, calculated using RCWA and reciprocity. (i) Spectral hemispherical emissivity of the patterned surface, which has the same unit cell as Celanovic \textit{et al.}~\cite{Celanovic2008} ($a=1\ \mu\textrm{m}$, $r=450\  \textrm{nm}$, $d=560\ \textrm{nm}$), compared to an unpatterned surface made of tungsten (blue). $\lambda_{c111}$ is the cutoff wavelength of the $\text{TE}_{111}$ mode supported by a cavity of radius $r$ and length $2d$.  (ii) Total hemispherical emissivity as a function of temperature. (d) Normalized $z$-component of the magnetic field $H_z$ in (i) the photonic crystal and (ii) a cylindrical cavity of radius $r$ and length $2d$. The similarities between them suggest that nanophotonic enhancement is partly caused by waveguiding.}
  \label{fig:1}
\end{figure*}

Consider a rotating anode X-ray tube, shown in Fig. \ref{fig:1}(a), under consideration as a canonical example of an X-ray tube. A tungsten cathode is heated to high temperatures ($>$2000\degree~C), generating an electron beam that is accelerated by the tube current and voltage, $I_t$ and $V_t$, and collides with the tungsten anode at the focal spot. Less than 1\% of the energy of the electron beam is converted into X-rays; the rest is converted into heat, which flows from the focal spot to the rest of the anode via conduction, then from the anode to the housing via radiation. The housing is cooled, e.g., by flowing air or oil over it. Typically, the anode is required to dissipate heat on the order of 10 kW, leading to high temperatures throughout the anode ($\sim$1000\degree~C), especially at the focal spot ($\sim$2500\degree~C). As previously mentioned, the result is a number of temperature-driven failure modes. These include anode melting, anode cracking due to thermal stress and fatigue (made worse by mechanical stress due to rotation at high speeds), anode evaporation and deposition on the interior surface of the housing, which blocks X-rays and can lead to the housing cracking (also known as crazing), and damage to seals resulting in loss of vacuum~\cite{Bushong2020failure,Behling2021failureI,Behling2021failureII}.

As previously mentioned, thermal radiation from the anode to the housing is an important mechanism of heat dissipation, but the emissivity of tungsten is relatively low ($\sim$0.1–0.4 temperature-dependent)~\cite{Minissale2017}. To address this limitation, we use photonic crystals to increase emissivity analogous to nanophotonic designs for thermophotovoltaics. This was theoretically predicted and experimentally demonstrated in a tungsten thermal emitter for use with thermophotovoltaic cells\cite{Celanovic2008,Yeng2012,Lenert2014,Chan2017,Sakakibara2022}. The concept is illustrated in Fig. \ref{fig:1}(a): by nanophotonically patterning the surface of the anode, e.g., with a square lattice of air holes as shown in Fig. \ref{fig:1}(a)(i), thermal radiation from the anode to the housing is enhanced. Such a design is realistic and can be fabricated, for example, using reactive ion etching~\cite{Celanovic2008,Yeng2012,Chan2017,Sakakibara2022}.

\subsection*{Nanophotonically-enhanced emissivity of the anode}\label{sec:nanophotonics}

We use rigorous coupled wave analysis (RCWA)~\cite{Jin2020} to simulate the optical response of a nanophotonic pattern identical to the one experimentally demonstrated by Celanovic \textit{et al.}~\cite{Celanovic2008} (period $a=1\ \mu\textrm{m}$, hole radius $r=450\ \textrm{nm}$, and hole depth $d=560\ \textrm{nm}$). This allows us to calculate the spectral directional absorptivity of the nanophotonic pattern, $\alpha(\lambda,\theta,\varphi)$, by simulating its interaction with plane wave launched toward it along the direction vector $\textbf{u}$. As a result of Kirchhoff’s law of thermal radiation (or Lorentz reciprocity), $\alpha(\lambda,\theta,\varphi)$ is equal to the spectral directional emissivity of the nanophotonic pattern, $\varepsilon(\lambda,\theta,\varphi)$~\cite{Kirchhoff1860original,SiegelHowellKirchhoff,Lorentz1896original,LandauLifshitzreciprocity}. This is illustrated in Fig. \ref{fig:1}(b). The application of Kirchhoff's law of thermal radiation is convenient because direct numerical simulations of thermal radiation are computationally expensive. They may involve, for example, Monte Carlo simulations of random dipoles or time-dependent modeling of a white noise source, both of which require a large ensemble average to observe a clear trend~\cite{Luo2004,Chan2006}. Calculating $\varepsilon(\lambda,\theta,\varphi)$ for many different values of $\lambda$, $\theta$, and $\varphi$ is necessary to calculate total hemispherical emissivity, to be discussed, but it can be prohibitively time-consuming without the computational convenience of Kirchhoff's law of thermal radiation.

Since the substrate is semi-infinite and lossy, $\alpha(\lambda,\theta,\varphi)=1-\rho(\lambda,\theta,\varphi)$, where $\rho(\lambda,\theta,\varphi)$ is the spectral directional reflectivity, encompassing all absorption, confinement, and scattering in the photonic crystal when it interacts with a plane wave. To understand the physics of nanophotonic enhancement, it is useful to look at the spectral hemispherical emissivity:

\begin{equation}\label{eq:spectralhemiemiss}
\varepsilon(\lambda)=\frac{1}{\pi}\int_{0}^{2\pi}\int_{0}^{\frac{\pi}{2}}[1-\rho(\lambda,\theta,\varphi)]\sin\theta\cos\theta\ d\theta\,d\varphi
\end{equation}

\noindent In the above equation, the angular dependence of $\varepsilon(\lambda,\theta,\phi)$ is integrated out, making it easier to see the role of resonances supported by the nanophotonic pattern, which we can interpret by observing the electromagnetic field at the resonant frequencies. The role of the nanophotonic pattern in tailoring the total hemispherical emissivity---which is the quantity typically used in thermal modeling---is elucidated by the following expression:

\begin{equation}\label{eq:totalhemiemiss}
\varepsilon(T)=\frac{\int_{0}^{\infty}\varepsilon(\lambda)B(\lambda,T)\,d\lambda}{\sigma T^4}.
\end{equation}

\noindent Essentially, Eq. (\ref{eq:totalhemiemiss}) states that the spectral ($\lambda$) and angular ($\theta$, $\varphi$) responses of the nanophotonic pattern, multiplied by the blackbody spectral radiance $B(\lambda,T)$, govern the temperature-dependent total hemispherical emissivity $\epsilon(T)$.

Figure \ref{fig:1}(c) demonstrates that nanophotonic patterning enhances the emissivity of the anode. In Fig.~\ref{fig:1}(c)(i), we observe an enhancement in the spectral hemispherical emissivity, $\varepsilon(\lambda)$, of a patterned anode (orange) compared to an unpatterned anode (blue). The patterned anode is more emissive in the spectral range of 1--3 \unit{\micro\metre}, which is desirable because it overlaps with the peak wavelength of the blackbody spectrum at typical anode temperatures (1000--2500 \unit{\degreeCelsius}). This is supported by Fig.~\ref{fig:1}(c)(ii), which shows that the temperature-dependent total hemispherical emissivity, $\varepsilon(T)$, is enhanced by a factor of approximately 2 even after integrating over all possible wavelengths and directions of emission.

Physically, the emissivity enhancement can be explained by recognizing that the design of the pattern is similar to a cylindrical cavity, which allows it to (1) efficiently outcouple light via waveguiding and (2) only outcouple light below the cutoff wavelength, allowing for spectral selectivity. We believe this is why $\varepsilon(\lambda)$ peaks close to $\lambda_{c111}=965$ \unit{\nano\metre}---the cutoff wavelength of the $\text{TE}_{111}$ mode supported by a cylindrical cavity of radius $r$ and length $2d$~\cite{Jackson}---and drops off at longer wavelengths. Figure~\ref{fig:1}(d) shows the real part of the $z$-component of the magnetic field, $\text{Re}(H_{z})$, at $\lambda_{c111}$, both in the pattern (i) and in a cylindrical cavity (ii). The similarity between the two cases supports our explanation of the emissivity enhancement and is consistent with Celanovic \textit{et al.}~\cite{Celanovic2008}. Furthermore, $\lambda_{c111}$ is close to the peak wavelengths corresponding to the temperatures at which $\varepsilon(T)$ is the highest (Fig. \ref{fig:1}(c)(i)), suggesting that the $\text{TE}_{111}$ mode is important for emissivity enhancement at high temperatures.

\subsection*{Enhanced performance enabled by nanophotonic thermal management}\label{sec:heattransfer}

\begin{figure*}
  \includegraphics[width=5.4in]{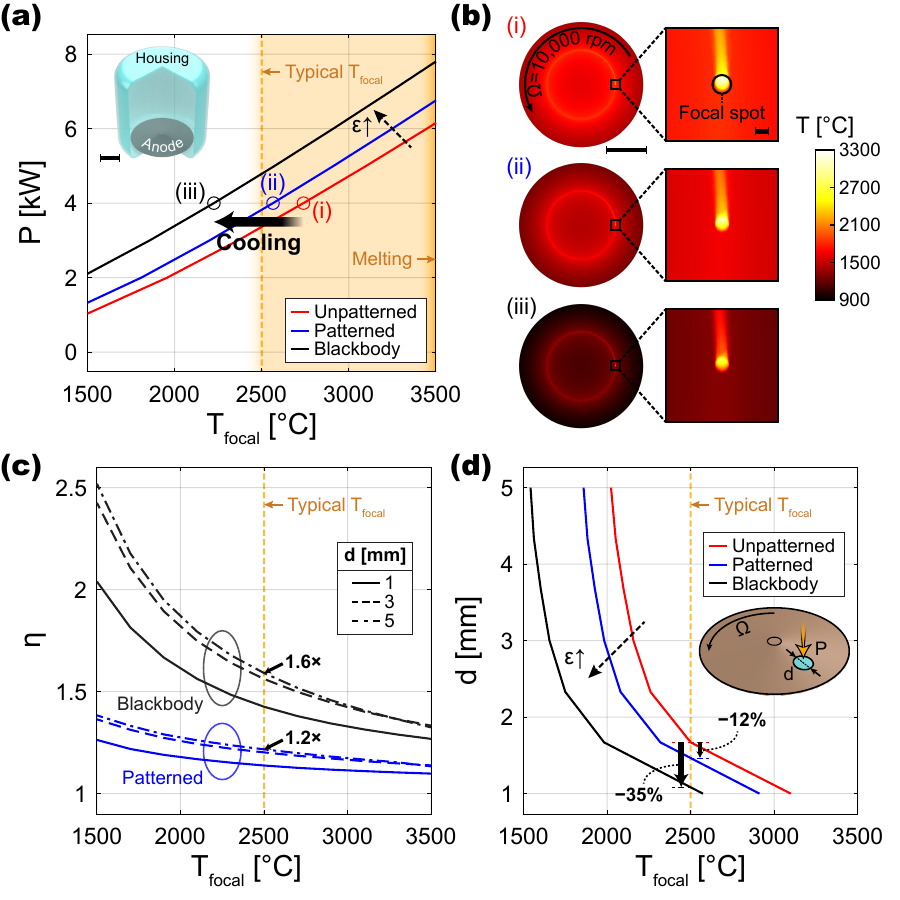}
  \caption{\textbf{Enhanced performance of an X-ray tube with a nanophotonically-patterned anode.} (a) Power input $P$ as a function of focal spot temperature $T_{\text{focal}}$ for focal spot size $d=1$ \unit{\milli\meter}. The red line corresponds to an unpatterned anode; blue, patterned; and black, blackbody. The dashed arrow indicates increasing emissivity, $\varepsilon$. The orange area indicates temperatures prone to common failure modes, up to the melting temperature of tungsten, approximately 3500 \unit{\celsius}. A render of the model geometry is shown inside the plot (scale bar = 5 \unit{\centi\meter}). (b) Anode temperature distribution at the points indicated by the open circles in (a), (i)--(iii) ($P=5$ \unit{\kilo\watt}, scale bar = 5 \unit{\centi\meter}). The zoomed-in images show the focal spot (scale bar = 1 \unit{\milli\meter}). (c) Power input enhancement compared to an unpatterned anode, $\eta \equiv P/P_{\text{unpatterned}}$, as a function of $T_{\text{focal}}$. The blue and black lines correspond to the patterned and blackbody anodes, respectively; different line styles correspond to different $d$-values. At $T_{\text{focal}}=2500$ \unit{\celsius}, indicated by the dashed orange line, $P$ can be enhanced by up to $1.2\times$ and $1.6\times$ in the patterned and blackbody cases, respectively. (d) $d$ as a function of $T_{\text{focal}}$. Smaller $d$-values translate to higher spatial resolution and coherence. For example, at $T_{\text{focal}}=2500$ \unit{\celsius}, $d$ can shrink by 12\% and 35\% in the patterned and blackbody cases, respectively. Inset: schematic showing $P$, $d$, and $\Omega$.}
  \label{fig:2}
\end{figure*}

To understand how nanophotonic thermal management can enhance the performance of the X-ray tube, we combine the results of our RCWA calculations with a finite element analysis software (COMSOL Multiphysics\textsuperscript{\textregistered}~\cite{comsol}) to develop a heat transfer model of the X-ray tube. We estimate the temperature distribution as a function of power input $P=I_{t}V_{t}$ and pay special attention to the focal spot temperature $T_{\text{focal}}$, which, to reiterate, is the hottest (i.e., limiting) temperature in the system. In this section, we will present the key results of our model; detailed descriptions including model geometries, equations, and software settings can be found in Sec. S3 of the Supporting Information.

Figure~\ref{fig:2}(a) shows how patterning the anode decreases the steady-state focal spot temperature $T_{\text{focal}}$ for a given input $P$ as a result of improved heat dissipation. In our model, rendered in the inset of Fig. \ref{fig:2}(a), the entire top surface of the anode except for the focal track is assumed to be patterned. The anode rotates at an angular frequency of $\Omega=10,000$ rpm and the housing is liquid-cooled by transformer oil. The design and dimensions of our model were informed by the typical characteristics of rotating anode X-ray tubes described in textbooks~\cite{Bushong2020design,Behling2021design}. As expected, increasing $\varepsilon$ shifts the $P$-$T_{\text{focal}}$ curves left in the direction of decreasing $T_{\text{focal}}$. This is because the anode is emitting more heat, thus cooling it down at a fixed $P$-value. Alternatively, if the X-ray tube operates at a fixed $T_{\text{focal}}$, patterning the anode would allow one to reach the desired set point while achieving a higher $P$-value. In other words, the X-ray tube can generate more X-rays without overheating above a safe temperature. Note that the black line in Fig. \ref{fig:2}(a) corresponds to the upper limit of $\varepsilon=1$, in which the anode is a blackbody.

The improved heat dissipation of the patterned anode is further supported by Figure \ref{fig:2}(b), which shows the temperature distribution on the surface of the anode for $P=5$ \unit{\kilo\watt}, corresponding to the points indicated by the open circles in Fig. \ref{fig:2}(a). The temperature distribution shows quantitative agreement with commonly accepted textbook values, in particular focal spot temperatures over and around 2500 \unit{\celsius} and bulk anode temperatures just above 1000 \unit{\celsius}~\cite{Behling2021temperature}. Qualitatively, the temperature distribution has the ``comet tail'' trailing behind the focal spot that is characteristic of rotating anode X-ray tubes~\cite{Behling2021design}. Figure~\ref{fig:2}(b) clearly illustrates how patterning the anode decreases the bulk anode temperature (i.e., the average temperature outside the relatively small focal spot and focal track) in addition to $T_{\text{focal}}$. Although $T_{\text{focal}}$ deserves attention because it is prone to causing catastrophic failure and thus more limiting, decreasing the bulk anode temperature can be beneficial as well. For example, it can lead to decreased thermal stress and anode evaporation.

Next, we define the power input enhancement relative to an unpatterned anode as

\begin{equation}\label{eq:powerenhancement}
    \eta \equiv \frac{P}{P_{\text{unpatterned}}}.
\end{equation}

\noindent Figure~\ref{fig:2}(c) shows $\eta$ as a function of $T_{\text{focal}}$ and $d$. Since the number of X-rays generated by the X-ray tube is directly proportional to $I_{t}V_{t}^{2}=PV_{t}$~\cite{Bushong2020xrays,Behling2021xrays}, $\eta$ is a metric of how much the dose can be enhanced as a result of nanophotonic patterning. Once again, the blue and black lines correspond to the cases in which the anode is patterned and a blackbody, respectively. Since the patterned surface area of the anode is $A_{\text{patterned}}=\pi(D-t\cot{\phi})(t\csc{\phi}-d)$ (where $D$, $t$, and $\phi$ are the anode diameter, thickness, and angle, respectively),\footnote[2]{This is nothing more than the curved surface area of a truncated cone of base diameter $D$, height $t$, and angle $\phi$ minus that of another truncated cone of slant height $d$ embedded in its center. A schematic of the anode can be found in Sec. S3 of the Supporting Information.} we expect $\eta$ to increase as $d$ decreases since the emitted power is proportional to $A_{\text{patterned}}$, as per the Stefan-Boltzmann law~\cite{SiegelHowellBlackbody}.

The reason $\eta$ decreases as $d$ shrinks is that heat transfer becomes conduction-limited. As $d$ decreases, the heat flux delivered to the focal spot, $4P/\pi d^2$, dramatically increases because of an inverse square law (where the focal spot is approximated as a circle of diameter $d$). The heat flux is approximately proportional to the surface temperature of the focal spot~\cite{Mills1995,Lienhard2024}. At the same time, the emitted power linearly increases as $d$ decreases (via the equation for $A_{\text{patterned}}$), and the power carried away by the rotation of the anode is proportional to $\dot{m}c_{p}T=\rho(\delta d)(\Omega D)c_{p}T$ (where $\dot{m}$ is the mass flow rate, and $\rho$ and $c_p$ are the density and specific heat of tungsten and $\delta$ is the penetration depth of the electron beam). Dimensional analysis of the heat equation tells us that conduction does not strongly depend on $d$ (its characteristic length is the radius of the anode, $D/2$). Thus, decreasing $d$ drives up the focal spot temperature while diminishing the effectiveness of advective cooling (via the rotation of the anode). Furthermore, before heat can be dissipated via thermal radiation, it is transported to the bulk anode via conduction, which does not explicitly depend on $d$. Thus, as a result of the conduction bottleneck, the improved heat dissipation provided by the nanophotonically-enhanced emissivity of the anode cannot keep up with the rapidly rising heat flux as $d$ shrinks. This is supported by scaling laws we derive in Sec. S5 in the Supporting Information, which are based on a thermal resistance model that does not directly account for the rotation of the anode and thermal energy storage due to the heat capacity of the anode.

In addition to enhancing the number of X-rays generated by the X-ray tube, nanophotonic thermal management has the potential to improve spatial resolution as well. This can be seen in Fig. \ref{fig:2}(d), which shows that as the anode becomes more emissive, the patterned and blackbody anodes can safely operate at a lower $d$-value for a desired set point. The thick black arrows in Fig. \ref{fig:2}(d) indicate that at $T_{\text{focal}}=2500$ \unit{\celsius}, for example, $d$ can be reduced by 12\% and 35\% in the cases of the patterned and a blackbody anode, respectively. As can be seen, attempting to operate an unpatterned anode at equivalent $d$-values would lead to temperatures in excess of 3000 \unit{\celsius}. The spatial resolution that can be achieved depends on $d$---more specifically, using the definition of geometric unsharpness (or penumbra) and the inverse proportionality between total unsharpness and spatial resolution $R$, it can be shown that

\begin{equation}\label{eq:spatialresolution}
    R\approx\frac{1}{\sqrt{(M-1)^{2}d^{2}+U_{\text{remaining}}^{2}}},
\end{equation}

\noindent where $M$ is the geometric magnification and $U_{\text{remaining}}$ are the remaining unsharpness terms, such as absorption, motion, and screen unsharpness~\cite{Christensen}. Since increasing $\varepsilon$ allows one to operate at lower $d$-values (Fig. \ref{fig:2}(d)), it can be argued that nanophotonic thermal management can lead to improved spatial resolution in X-ray imaging (Eq. (\ref{eq:spatialresolution})).

It can also be argued that nanophotonic thermal management can help with the spatial coherence of X-ray tubes. The coherence length of an X-ray source is given by

\begin{equation}
    \ell_{c} = \frac{\Lambda z}{d},
\end{equation}

\noindent where $\Lambda$ is the wavelength of X-rays, $z$ is the distance from the focal spot, and $d$ is the focal spot size as before~\cite{Chen2010}. Typically, a longer $\ell_c$ corresponds to a more spatially coherent X-ray source. In the examples shown in Fig. \ref{fig:2}, where $d$ decreases by 12\% and 35\% in the patterned and blackbody cases, the corresponding increase in $\ell_c$ is by 1.1$\times$ and 1.5$\times$, respectively. At lower values of $T_{\text{focal}}$, the reduction in $d$ and thus increase in $\ell_c$ can be much more dramatic. Furthermore, a longer $\ell_c$ enabled by nanophotonic thermal management can help relax the conditions required for PCI in the Talbot-Lau configuration, in which the gratings used to interfere X-rays are required to have a period on the order of $\ell_c$. Therefore, an increase in $\ell_c$ means the period of the gratings can be longer, making them easier to fabricate.

\subsection*{Practical aspects of tungsten photonic crystals}\label{sec:practical}

Using strategies from the field of thermophotovoltaics, where tungsten photonic crystals are heated to high temperatures to be used as thermal emitters~\cite{Celanovic2008,Yeng2012,Lenert2014,Chan2017,Sakakibara2022}, we predict that nanophotonically patterning the anode of a rotating anode X-ray tube---which is prone to failure modes of thermal origin~\cite{Bushong2020failure,Behling2021failureI,Behling2021failureII}---can lead to improvements in thermal management. In turn, we predict that nanophotonic thermal management can lead to enhancements in dose and spatial resolution, both of which are critical to enabling transformative imaging modalities such as PCI.

Throughout this work, we also modeled what would happen in the limiting case that the anode is a blackbody. This can, in fact, be achieved in practice, though not necessarily through the use of nanophotonic patterning in the usual sense (i.e., photonic crystals). Prior works have shown that it is possible to increase the emissivity of refractory metals such as tungsten above 99\% by depositing vertically aligned carbon nanotube arrays (VANTAs), which are said to be stable \textit{in vacuo} at high temperatures~\cite{Lenert2014}, or by surface texturing using laser ablation~\cite{park2024I,park2024II,Verma2025}.

It is worth noting that patterning the anode does not significantly affect X-ray generation. In Sec. S6 of the Supporting Information, we show the spectrum and angular distribution of X-rays generated via bremsstrahlung as a function of the energy angle of incidence of the electron beam (a proxy for the anode angle), calculated using Geant4~\cite{Agostinelli2003geant4}. There are no notable differences between the spectrum and angular distribution of X-rays generated by a patterned anode compared to an unpatterned anode. This can be explained by the incongruent length scales of the pattern (on the order of 1 \unit{\micro\meter}) and the generated X-rays (sub-\unit{\nano\meter}), which differ by at least 3--4 orders of magnitude.

A potential drawback of patterning the anode is increasing the evaporation rate of tungsten since the pattern increases the surface area of the anode. However, there are strategies to mitigate this. For example, the air holes in the pattern studied in this work can be filled with hafnia, which is an inert dielectric material in the mid-infrared spectrum that acts as a passivation layer and improves thermochemical stability~\cite{Chan2017,Sakakibara2022}. In the case of a blackbody anode, which can be achieved by depositing VANTAs on the surface of the anode, the coating's stability can become a question under vacuum conditions at high temperatures~\cite{Lenert2014}. Refractory metals that have been laser-ablated to increase their emissivity above 99\% have surface oxidation, which could affect their thermochemical stability \textit{in vacuo}, although to what extent is unknown~\cite{Verma2025}.

\section*{Conclusion}\label{sec:conclusion}

In summary, we predict that by nanophotonically enhancing the emissivity of the anode, we can  more effectively cool it at a fixed power input or, inversely, enhance the power input---which is proportional to the number of X-rays generated---at a fixed focal spot temperature. In principle, we can also decrease the focal spot size at a fixed power input, which can improve the spatial resolution of X-ray imaging as well as the spatial coherence of X-rays, which is important for imaging modalities such as phase-contrast imaging. This approach can be extended to other X-ray tube designs (e.g., stationary anode) and anodes made of different materials (e.g., molybdenum and graphite). Our results suggest that nanophotonic thermal management can be a path toward using X-ray tubes in high-dose imaging modalities, with high spatial resolution.

\section*{Methods}

\subsection*{Nanophotonic model}

We used the RCWA software grcwa (version 0.1.2)~\cite{Jin2020} to nanophotonically model the pattern described in this work~\cite{Celanovic2008}. To summarize the details in the main text and Supporting Information Sec. S2: we modeled a unit cell consisting of a semi-infinite slab of tungsten (relative permittivity given by a Drude-Lorentz model~\cite{Rakić1998}) with an air hole of radius $r=450$ \unit{\nano\meter} and depth $d=560$ \unit{\nano\meter}. The period of the pattern was $a=1$ \unit{\micro\meter}. By launching a plane wave at the pattern and invoking Kirchhoff's law of thermal radiation~\cite{Kirchhoff1860original,SiegelHowellKirchhoff}, RCWA was used to calculate the total hemispherical emissivity via Eqs. (\ref{eq:spectralhemiemiss}) and (\ref{eq:totalhemiemiss}), which was an input to the heat transfer model.

\subsection*{Heat transfer model}

We used COMSOL Multiphysics\textsuperscript{\textregistered}~\cite{comsol} (version 5.5, with the ``Heat Transfer with Surface-to-Surface Radiation Module'') to calculate the temperature distribution in the X-ray tube using the finite element method. The model geometry included the anode and the housing but not the filament because it is a relatively small, self-viewing surface and should not significantly contribute to radiative heat transfer. The anode was allowed to rotate at an angular frequency of $\Omega$, and the electron beam bombarding the anode was modeled as a heat flux $P=I_{t}V_{t}$ entering the focal spot. The exterior surface of the housing and the bottom surface of the stem were convectively cooled, while the remaining surfaces were subject to radiation boundary conditions, provided the emissivity of each surface. Only the top of the anode excluding the focal track was assumed to be nanophotonically patterned, meaning the total hemispherical emissivity of patterned tungsten, calculated using RCWA, was assigned to this surface, while all other tungsten surfaces were treated as unpatterned. The details of our heat transfer model, including the model geometry and its dimensions, equations describing the boundary conditions, material properties, and mesh can be found in Sec. S3 of the Supporting Information.

\section*{Acknowledgement}

The authors thank C. Wilson and L.-E. Cole for helpful discussions on heat transfer modeling; S. Verma for helpful discussions on laser-ablated refractory metals and their thermochemical stability under vacuum conditions; and W. Michaels for helpful discussions on focal spot size and spatial coherence. This work was supported in part by DARPA Agreement Number HO0011249049, as well as in part by the U.S. Army Research Office through the Institute for Soldier Nanotechnologies at MIT, under Collaborative Agreement Number W911NF-23-2-0121. S. Pajovic gratefully acknowledges support from the NSF GRFP under Grant Number 2141064 and the MathWorks Engineering Fellowship. C. Roques-Carmes is supported by a Stanford Science Fellowship. S. Choi acknowledges support from the Korea Foundation for Advanced Studies Overseas PhD Scholarship.

\section*{Supporting Information Available}

Additional literature review on thermal management needs of different imaging modalities and prior work on thermal management in X-ray tubes; details of the nanophotonic model; details of the finite element method heat transfer model using COMSOL Multiphysics\textsuperscript{\textregistered}; additional results investigating other design parameters of rotating anode X-ray tubes; thermal resistance heat transfer model and scaling laws; and simulated spectral and angular distribution of X-rays emitted by the nanophotonically-patterned anode using Geant4.

\bibliographystyle{ieeetr}
\bibliography{references} 
 

\newpage
\include{supplementary_arXiv}

\end{document}

%% file: supplementary_arXiv.tex
\clearpage
\onecolumngrid

\begin{center}

\textbf{\Large Supporting Information for:\\Nanophotonic thermal management in X-ray tubes}
\linebreak[4]

Simo~Pajovic$^{1,*}$, Charles~Roques-Carmes$^{2,3}$, Seou~Choi$^{2}$, Steven~E.~Kooi$^{4}$, Rajiv~Gupta$^{5}$, Michael~E.~Zalis$^{5}$, Ivan~Čelanović$^{4}$, Marin~Soljačić$^{6}$
\linebreak[4]

$^{1}$\textit{Department of Mechanical Engineering, Massachusetts Institute of Technology, Cambridge, MA, 02139, U.S.A.}\\
\vspace{0.06cm}$^{2}$\textit{Research Laboratory of Electronics, Massachusetts Institute of Technology, Cambridge, MA, 02139, U.S.A.}\\
\vspace{0.06cm}$^{3}$\textit{E. L. Ginzton Laboratory, Stanford University, Stanford, CA, 94305, U.S.A.}\\
\vspace{0.06cm}$^{4}$\textit{Institute for Soldier Nanotechnologies, Massachusetts Institute of Technology, Cambridge, MA, 02139, U.S.A.}\\
\vspace{0.06cm}$^{5}$\textit{Department of Radiology, Massachusetts General Hospital, Boston, MA, 02114, U.S.A.}\\
\vspace{0.06cm}$^{6}$\textit{Department of Physics, Massachusetts Institute of Technology, Cambridge, MA, 02139, U.S.A.}

\end{center}

\setcounter{equation}{0}
\setcounter{figure}{0}
\setcounter{table}{0}
\setcounter{page}{1}
\pagenumbering{Roman}
\setcounter{section}{0}
\makeatletter
\renewcommand{\thesection}{S\arabic{section}}
\renewcommand{\theequation}{S\arabic{equation}}
\renewcommand{\thefigure}{S\arabic{figure}}
\renewcommand{\thetable}{S\arabic{table}}


\section{Challenges and approaches to thermal management in X-ray tubes}\label{sec:review}

In this section, we further discuss the thermal management needs of different imaging modalities and summarize the key milestones in the development of thermal management in X-ray tubes. To summarize the discussion in the main text: a major issue with generating X-rays via bremsstrahlung is that much less than 1\% of the power input (on the order of 10 \unit{\kilo\watt} or less) is converted into X-rays, and the rest is converted into heat. In X-ray tubes, this results in temperature-driven failure modes such as melting of the anode, cracking due to thermal stress and fatigue, evaporation and subsequent deposition of tungsten on the interior surface of the housing (which can block X-rays and cause arcing, which in turn can crack the housing, also known as crazing), and loss of vacuum via damaged seals~\cite{Bushong2020failure,Behling2021failureI,Behling2021failureII}. The existence of these failure modes of X-ray tubes prevents operation at high power and small focal spot sizes---rendering some imaging modalities such as phase-contrast imaging (PCI) infeasible in clinical and industrial settings---and has driven the development of highly sophisticated thermal management solutions, to be discussed.

The problem of thermal management in X-ray tubes is further complicated by the fact that different imaging modalities face unique challenges. Computed tomography (CT) is demanding in almost every aspect: X-ray photon flux (brightness), photons per patient (dose), and even the photon energies used for imaging (related to the tube voltage $V_t$). Mammography demands high spatial resolution at low $V_t$, which can be challenging because of space-charge-limited current. Surgical applications demand a small but continuous photon flux over the course of an operation~\cite{Behling2021design}. PCI is of particular interest~\cite{Momose2005review,Bravin2012review} because it does not rely on the absorption of X-rays, meaning it can be more sensitive to soft tissues (which are not very dense and absorptive) and still work using lower doses, which can be safer for patients. However, PCI demands both high brightness and high spatial coherence, meaning the path differences between emitted photons need to be commensurate with their wavelengths (on the order of \num{e-11}~\unit{\metre})~\cite{Bartzsch2017}. For example, propagation-based PCI requires a 10~\unit{\micro\metre} focal spot size~\cite{Behling2021phase}, but as previously mentioned, this is limited by thermal management. For this reason, the majority of PCI experiments, especially seminal works, used synchrotron radiation~\cite{Momose1995phasesynch,Momose1996phasesynch,Momose2003phasesynch,Westneat2003phasesynch,Weitkamp2005phasesynch,Zhu2010phasesynch,Wen2013phasesynch}, although there are approaches to PCI that can work using specially designed high-brightness X-ray tubes~\cite{Pfeiffer2006phasetube,Olivo2007phasetube,Pfeiffer2008phasetube,Donath2010phasetube,Munro2012phasetube,Tapfer2012phasetube,Miao2015phasetube,Miao2016phasetube}.

With these challenges in mind, the need for high power, high spatial resolution, and cost reduction (including capital cost in the form of the anode, the most expensive part of the X-ray tube, and running cost in the form of energy consumption) has driven the development of X-ray tube designs that are better at dissipating heat~\cite{Behling2021design,Iversen1984}. Stationary anode X-ray tubes as well as miniature X-ray tubes liquid cool the anode, but as a result of their size, the emitted power is 1--2 orders of magnitude smaller than, e.g., a rotating anode X-ray tube~\cite{Behling2021stationary,Nadella2015}. Rotating anode X-ray tubes are the most common design, striking a balance between brightness and spatial resolution. The innovation is rotation at high speeds in excess of thousands of rpm to distribute heat around the circumference of the anode~\cite{Behling2021design,Bushong2020design}. Liquid-cooled rotating anode X-ray tubes also exist~\cite{Iversen1984} but are not widely used. A more recent design, the rotating envelope X-ray tube, represents a breakthrough in thermal management~\cite{Schardt2004}. The bulk anode temperature is not more than a few hundred degrees Celsius, and the design is compact, cost-effective, and has a power rating comparable to its predecessors. The challenges of rotating envelope X-ray tubes include overcoming fluid resistance and cavitation, thermal stress due to large temperature gradients, and inefficient conversion of electrons to X-rays due to electron backscattering~\cite{Behling2021design,Schardt2004}. The importance of controlling electron backscattering has been acknowledged~\cite{Davies1959}, as backscattered electrons can heat up other parts of the X-ray tube. Another relatively recent innovation is the liquid metal jet X-ray tube, which is brighter than rotating anode X-ray tubes by two orders of magnitude. The high speed of the liquid metal jet accounts for one order of magnitude, and the ability to tolerate a higher thermal load accounts for another. Tin or gallium are used as the anode material instead of tungsten because of their relatively low melting point and bright K-line emission~\cite{Hemberg2003}. Behling \textit{et al.} proposed a similar design that uses a jet of microparticles instead of liquid metal~\cite{Behling2024}. Other approaches have included using both rotation and reciprocating motion to more evenly distribute heat~\cite{Bartzsch2017}, embedding a heat pipe in the anode~\cite{Trubitsyn2019}, and thermionic cooling~\cite{Chen2024}.


\section{Nanophotonic model of the anode}\label{sec:nanophotonic}

We used the RCWA software grcwa (version 0.1.2)~\cite{Jin2020} to model the interaction between the nanophotonic pattern of interest~\cite{Celanovic2008} and plane waves of different wavelengths $\lambda$ launched from different polar and azimuthal angles of incidence $\theta$ and $\varphi$. As discussed in the main text, this allowed us to calculate the spectral directional reflectivity $\rho(\lambda,\theta,\varphi)$ and, in turn, the spectral directional absorptivity $\alpha(\lambda,\theta,\varphi)$. Then, via Kirchhoff's law of thermal radiation~\cite{Kirchhoff1860original,SiegelHowellKirchhoff}, $\alpha(\lambda,\theta,\varphi)$ is equal to the spectral directional emissivity $\varepsilon(\lambda,\theta,\varphi)$, which we used to calculate the total hemispherical emissivity $\varepsilon(T)$ via Eq. (2) in the main text. This was used as an input to the heat transfer model.

\begin{figure}[h!]
    \centering
    \includegraphics{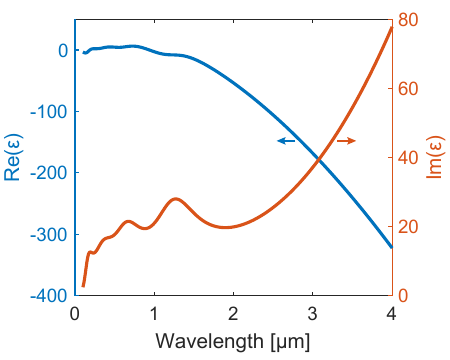}
    \vspace{-0.3cm}
    \caption{Dielectric function of tungsten, calculated using a Drude-Lorentz model fitted to experimental data from~\cite{Rakić1998}. The real and imaginary parts of the dielectric function ($\text{Re}(\varepsilon)$ and $\text{Im}(\varepsilon)$) are plotted against the left blue and right orange vertical axes, respectively, as indicated by the arrows.}
    \label{fig:wdielectric}
\end{figure}

The model geometry was a single unit cell of the nanophotonic pattern, with periodic boundary conditions in the lateral directions and semi-infinite in the transverse direction to represent a slab of tungsten (the anode). The plane waves were launched from air. We assumed the relative permittivity of tungsten is given by a Drude-Lorentz model fitted to experimental data~\cite{Rakić1998}, plotted in Fig. \ref{fig:wdielectric}. It can be temperature-dependent~\cite{Minissale2017}, but in this work, we assumed the dielectric function is constant with respect to temperature. We also assumed the relative permeability of tungsten is 1.

We calculated $\varepsilon(\lambda,\theta,\varphi)$ for $\lambda$-values ranging from 0.6 \unit{\micro\meter} to 100 \unit{\micro\meter}, which was based on Wien's law~\cite{Wien1896original,SiegelHowellBlackbody} and the predicted temperature range of the anode (from textbooks~\cite{Bushong2020design,Behling2021design} and our thermal resistance model, described in Sec. S4 of the Supporting Information). Since the unit cell has C4v symmetry, we only needed to calculate $\varepsilon(\lambda,\theta,\varphi)$ for $\varphi$-values ranging from 0 to $\pi/2$. This meant that the integration with respect to $\varphi$ in Eq. (1) in the main text could be carried out from 0 to $\pi/2$ (instead of 0 to $2\pi$) and multiplied by 4, i.e., $\int_{0}^{2\pi}d\varphi \rightarrow 4\int_{0}^{\pi/2}d\varphi$.

In the case of the unpatterned anode, we used the Fresnel equations for planar interfaces to calculate $\rho(\lambda,\theta,\varphi)$. More specifically, if the reflection coefficients are given by

\begin{align}
    r_s &= \frac{k_{z,2} - k_{z,1}}{k_{z,2} + k_{z,1}},\label{eq:rs}\\
    r_p &= \frac{\epsilon_{1}k_{z,2} - \epsilon_{2}k_{z,1}}{\epsilon_{1}k_{z,2} + \epsilon_{2}k_{z,1}},\label{eq:rp}
\end{align}

\noindent then $\rho(\lambda,\theta,\varphi) = (|r_s|^2 + |r_p|^2)/2$, where $\epsilon_i$ is the relative permittivity of medium $i$ (air or tungsten), $k_{z,i} = \sqrt{\epsilon_{i}k_{0}^{2}-k_{x}^{2}}$ is the $z$-component of the wavevector in medium $i$, $k_0 = 2\pi/\lambda$ is the wavenumber, and $k_x = k_{0}\sin\theta$ is the $x$-component of the wavevector, which is conserved in both media~\cite{LandauLifshitzreflection,BornWolf}. Note that Eqs. (\ref{eq:rs}) and (\ref{eq:rp}) do not depend on $\varphi$ since a planar interface between two isotropic media has continuous translational symmetry. This meant that the integration with respect to $\varphi$ in Eq. (1) in the main text evaluated to $2\pi$ ($\int_{0}^{2\pi}d\varphi \rightarrow 2\pi$).


\section{Finite element method model of the X-ray tube}\label{sec:fem}

\subsection{Model geometry}\label{sec:geometry}

\begin{figure}[h!]
    \centering
    \includegraphics[width=4.5in]{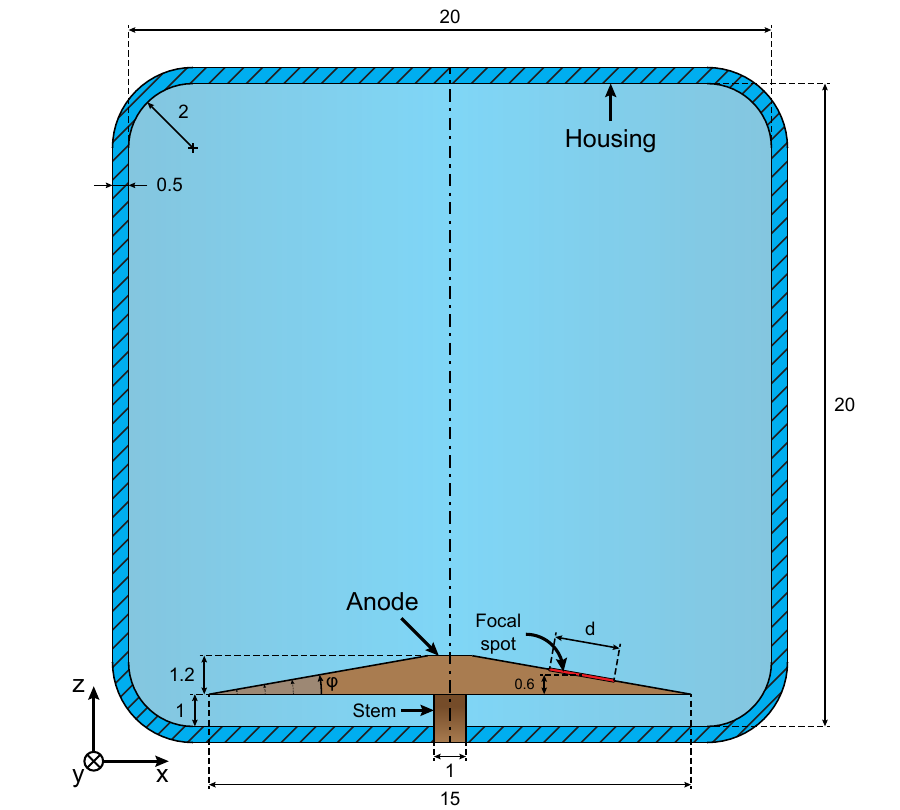}
    \vspace{-0.3cm}
    \caption{Side view, cross-sectional scale drawing of the geometry used in our FEM model. Dimensions are in units of \unit{\centi\meter}. The focal spot, which is located exactly halfway up the anode, is indicated by the thin strip of red. Excluding the focal spot, the geometry has cylindrical symmetry about the dash-dotted line. The focal spot size $d$ and anode (interior) angle $\varphi$ were variable.}
    \label{fig:sideviewfem}
\end{figure}

\begin{figure}[h!]
    \centering
    \includegraphics[width=4.5in]{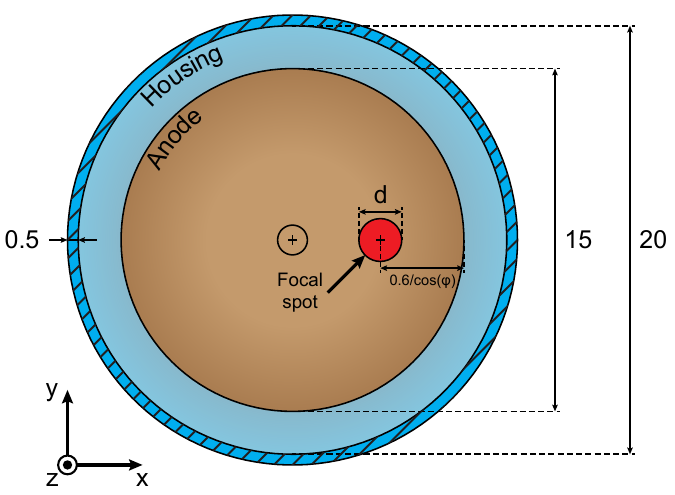}
    \vspace{-0.3cm}
    \caption{Top view, cross-sectional scale drawing of the geometry used in our FEM model. Dimensions are in units of \unit{\centi\meter}. The lateral position of the focal spot depends on the anode angle $\varphi$, as shown (0.6 \unit{\centi\meter} is half the thickness of the anode).}
    \label{fig:topviewfem}
\end{figure}

We used COMSOL Multiphysics\textsuperscript{\textregistered}~to model the X-ray tube using the finite element method (FEM). Figures \ref{fig:sideviewfem} and \ref{fig:topviewfem} show the dimensions in \unit{\centi\meter} of the geometry used in our FEM model of the X-ray tube. The model geometry was not based on a particular design or commercial product but rather images and typical dimensions found in textbooks~\cite{Bushong2020design,Behling2021design}. One part excluded from the model geometry is the cathode, which includes a tungsten filament and an assembly holding it in place and connecting it to the power supply at $V_{t}$ and $I_{t}$. This simplification is justified as follows:

\begin{enumerate}
    \item \textbf{Filament.} The filament, which thermionically emits the electrons which collide with the anode, is one of the hottest parts of an X-ray tube, reaching approximately 2000--2500 \unit{\celsius}~\cite{Behling2021design}. In spite of this, the filament is shaped like a relatively small (1 \unit{\centi\meter} long by 1 \unit{\milli\meter} wide~\cite{Bushong2020design}) coil, which is self-viewing. Depending on the geometry of the coil, the effective emissivity of the filament can be 30--60\% lower than the emissivity of tungsten as a result of the ``shadow factor''~\cite{Raczkevi1988}, or view factor between the filament and itself. This is compounded by the fact that the emissivity of tungsten is 0.2--0.4, which is relatively low in the first place. Since the filament has a low emissivity and is relatively small and self-viewing, we can conclude that its contribution to radiative heat transfer is negligible.
    \item \textbf{Cathode assembly.} Typically, the cathode assembly and electronics are covered by a ``shroud''~\cite{Bushong2020design}, which is metallic and therefore has a low emissivity. In particular, the cup is made of a highly reflective material to help thermally isolate the cathode and prevent it from losing heat.
\end{enumerate}

\newpage
\subsection{Boundary conditions}\label{sec:bcs}

\begin{figure}[h!]
    \centering
    \includegraphics[width=4.5in]{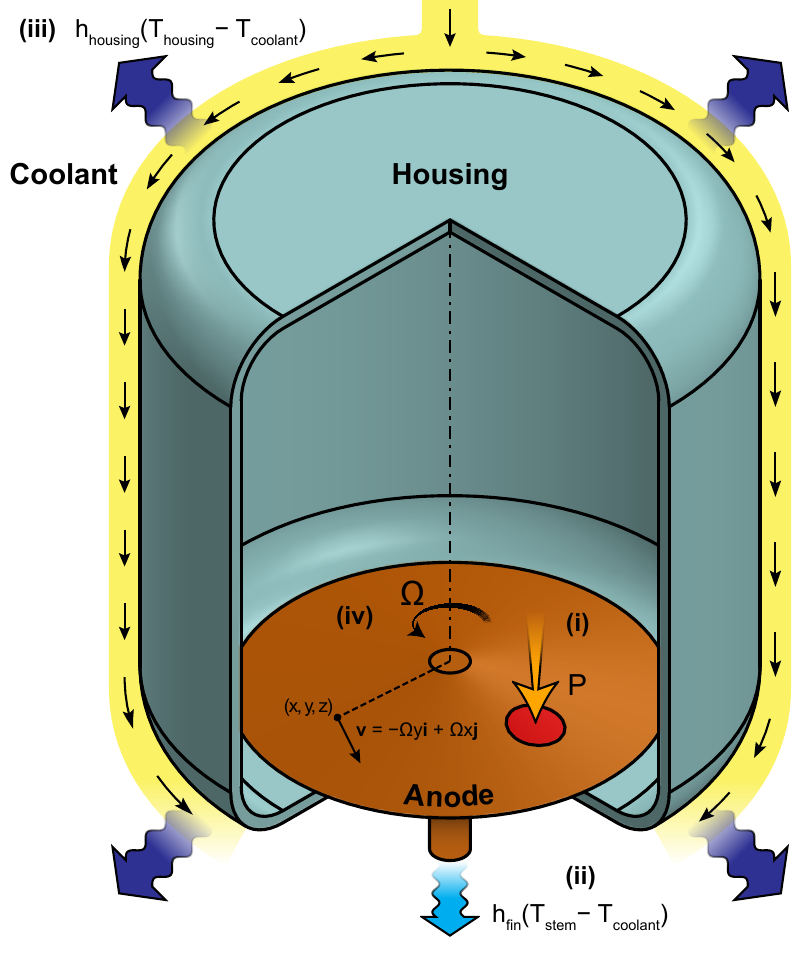}
    \vspace{-0.45cm}
    \caption{Schematic of the boundary conditions in our FEM model, except for radiation, which is applied to all surfaces within the housing. (i) Heat flux boundary condition. Power input $P$ (equal to $I_{t}V_{t}$, the product of the tube current and voltage) enters the focal spot, whose area is approximately $\pi d^{2}/4$ (neglecting the curvature of the anode). (ii) Convection from the stem. The coolant at temperature $T_{\text{coolant}}$ removes heat from the part of the stem sticking out of the housing, treated as a fin at temperature $T_{\text{stem}}$. The equation for the effective heat transfer coefficient, $h_{\text{fin}}$, is described in Sec. \ref{sec:bcs}. (iii) Convection from the housing. The coolant removes heat from the housing, which is at $T_{\text{housing}}$ and absorbs thermal radiation from the anode. The heat transfer coefficient $h_{\text{housing}}$ is calculated using heat transfer correlations for flow over a flat plate, also described in Sec. \ref{sec:bcs}. (iv) Rotation at an angular frequency of $\Omega$. The velocity field of the anode and stem was $\textbf{v}=-\Omega y\textbf{i}+\Omega x\textbf{j}$, where $\textbf{i}$ and $\textbf{j}$ are unit vectors in the $x$- and $y$-directions.}
    \label{fig:bcs}
\end{figure}

The ``Heat Transfer with Surface-to-Surface Radiation'' multiphysics coupling in COMSOL Multiphysics\textsuperscript{\textregistered}~accounts for radiation at boundaries by adding a source of the form

\begin{equation}\label{eq:radbc}
    q_{\text{rad}} = \varepsilon[G-e_{b}(T)]
\end{equation}

\noindent to the heat equation at boundaries, where $\varepsilon$ is the emissivity, $G$ is the irradiation, and $e_{b}(T)$ is the intensity of a blackbody at temperature $T$~\cite{comsolmanual}. This boundary condition is applied to all surfaces in the model geometry, except the exterior surface of the housing (to be discussed). There are three additional boundary conditions we apply, which are illustrated in Fig. \ref{fig:bcs}.

\textit{Figure \ref{fig:bcs}(i)}---A heat flux boundary condition is applied to the focal spot. This models the heating of the focal spot as a result of the $<1$\% of the power input $P$ being converted into X-rays.

\textit{Figure \ref{fig:bcs}(ii)}---A convection boundary condition is applied to the bottom of the stem. Since part of the stem sticks out of the housing, we model it as an infinitely long fin with an effective heat transfer coefficient that captures its heat exchange with the coolant~\cite{Mills1995,Lienhard2024}. In particular, it can be shown that

\begin{equation}\label{eq:finbc}
    q_{\text{fin}} = h_{\text{fin}}(T_{\text{stem}} - T_{\text{coolant}}),
\end{equation}

\noindent where $h_{\text{fin}}=2\sqrt{h_{\text{stem}}k_{\text{W}}/D_{\text{stem}}}$ is the heat transfer coefficient representing convection from the stem to the coolant, in which $k_{\text{W}}$ is the thermal conductivity of tungsten and $D_{\text{stem}}$ is the stem diameter, and $T_{\text{stem}}$ and $T_{\text{coolant}}$ are the temperatures of the stem and coolant, respectively. To calculate $h_{\text{stem}}$, we treat the stem as a cylinder in crossflow and use the following heat transfer correlation:

\begin{equation}\label{eq:nustem}
    \overline{\text{Nu}}_D = 0.3 + \frac{0.62\text{Re}_{D}^{\frac{1}{2}}\text{Pr}^{\frac{1}{3}}}{\left[1+\left(\frac{0.4}{\text{Pr}}\right)^{\frac{2}{3}}\right]^{\frac{1}{4}}},
\end{equation}

\noindent where $\overline{\text{Nu}}_{D}=h_{\text{stem}}D/k_{\text{oil}}$ is the average Nusselt number, $\text{Re}_{D}=U_{\infty}D/\nu_{\text{oil}}$ is the Reynolds number, and $\text{Pr}=\nu_{\text{oil}}/\alpha_{\text{oil}}$ is the Prandtl number~\cite{Mills1995,Churchill1977}. In addition, $k_{\text{oil}}$, $\nu_{\text{oil}}$, and $\alpha_{\text{oil}}$ are the thermal conductivity, kinematic viscosity, and thermal diffusivity of the coolant (in this case, transformer oil), which can be temperature-dependent~\cite{Nadolny2017,Kurzweil2021}. $U_{\infty}$ is the velocity of the coolant, assumed to be 1 \unit{\meter\per\second} in most of our calculations. Equation (\ref{eq:nustem}) is a valid approximation provided that $\text{Re}_{D}<4000$~\cite{Mills1995}. Alternatively, we can treat that stem as a cylinder rotating in a stationary fluid and use a heat transfer correlation such as Eq. (4.137) in Mills' textbook~\cite{Mills1995}. However, the resulting value of $h_{\text{stem}}$ is not much higher or lower (i.e., by less than an order of magnitude) than the one provided by Eq. (\ref{eq:nustem}).

\textit{Figure \ref{fig:bcs}(iii)}---A convection boundary condition is applied to the exterior surface of the housing, which can be written as 

\begin{equation}\label{eq:housingbc}
    q_{\text{housing}}=h_{\text{housing}}(T_{\text{housing}}-T_{\text{coolant}}).
\end{equation}

\noindent Again, $h_{\text{housing}}$ is the heat transfer coefficient representing convection from the housing to the coolant and $T_{\text{housing}}$ is the temperature of the housing. In theory, $h_{\text{coolant}}$ can be as large as we want because we can arbitrarily increase the coolant velocity, but in practice, it can be limited by the volumetric flow rate of commercially available pumps, the geometry of the X-ray tube, and more. Ultimately, for the sake of concreteness, we use the following heat transfer correlation to calculate $h_{\text{housing}}$:

\begin{equation}
    \overline{\text{Nu}}_L = 0.664\text{Re}_{L}^{1/2}\text{Pr}^{1/3},
\end{equation}

\noindent where $\overline{\text{Nu}}_L = h_{\text{housing}}L/k_{\text{oil}}$ and $L$ is a characteristic length equal to the average chord length $2D_{\text{housing}}/\pi$ plus the height of the housing, $H_{\text{housing}}$. This approach treats the housing like it has been ``unrolled'' into a flat plate~\cite{Mills1995}. Equation (\ref{eq:housingbc}) overrides the radiation boundary condition, Eq. (\ref{eq:radbc}).

Finally, although it is not a boundary condition per se, we set the velocity field of the anode and stem as

\begin{equation}\label{eq:velocityfield}
    \textbf{v}=-\Omega y\textbf{i}+\Omega x\textbf{j},
\end{equation}

\noindent where $\textbf{i}$ and $\textbf{j}$ are unit vectors in the $x$- and $y$-directions. This is illustrated in Fig. \ref{fig:bcs}(iv) and models the rotation of the anode.

\subsection{Material properties}\label{sec:materials}

\begin{figure}[h!]
    \centering
    \includegraphics[width=\textwidth]{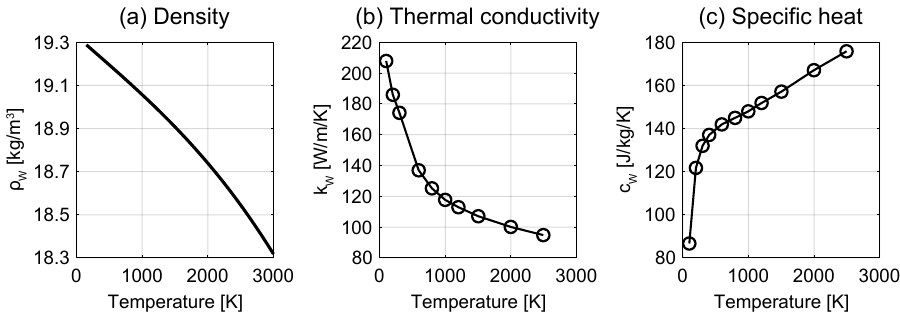}
    \vspace{-0.6cm}
    \caption{Thermophysical properties of tungsten used in this work. (a) Density $\rho_{\text{W}}$ was calculated using the White-Minges fit~\cite{WhiteMinges,Tolias2017}. (b) Thermal conductivity $k_{\text{W}}$ and (c) specific heat $c_{\text{W}}$ were taken from~\cite{Incropera2007} and~\cite{Smith1991}. The open circles indicate the actual data points extracted using WebPlotDigitizer~\cite{WebPlotDigitizer}.}
    \label{fig:wthermophys}
\end{figure}

\begin{figure}[h!]
    \centering
    \includegraphics{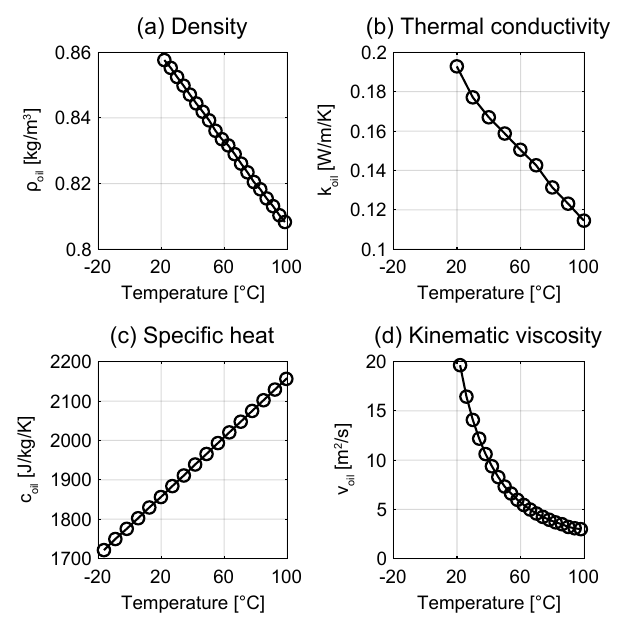}
    \vspace{-0.6cm}
    \caption{Thermophysical properties of transformer oil. (a) Density $\rho_{\text{oil}}$, (b) thermal conductivity $k_{\text{oil}}$, and (c) specific heat $c_{\text{oil}}$ were taken from~\cite{Kurzweil2021}. (d) Kinematic viscosity $\nu_{\text{oil}}$ was taken from~\cite{Linseis}. Again, the open circles indicate the actual data points extracted using WebPlotDigitizer~\cite{WebPlotDigitizer}.}
    \label{fig:oilthermophys}
\end{figure}

Unless otherwise stated, we extracted the material properties from plots in the literature using WebPlotDigitizer~\cite{WebPlotDigitizer}. The tabulated material property data were linearly interpolated by COMSOL Multiphysics\textsuperscript{\textregistered}. When the data needed to be extrapolated, we assumed it was constant outside the data range. The thermal conductivity and specific heat of tungsten were taken from~\cite{Incropera2007} and~\cite{Smith1991}, while the density of tungsten in \unit{\gram\per\cubic\centi\meter} is given by the White-Minges fit, which is empirical~\cite{WhiteMinges,Tolias2017}. These are plotted in Fig. \ref{fig:wthermophys}. The coolant was transformer oil, whose thermophysical properties are plotted in Fig. \ref{fig:oilthermophys}. The density, specific heat, and kinematic viscosity of transformer oil were taken from~\cite{Kurzweil2021}. The thermal conductivity was taken from a plot provided by LINSEIS Inc.~\cite{Linseis}. The density, specific heat, and thermal conductivity of glass were 2.23 \unit{\gram\per\cubic\centi\meter}, 820 \unit{\joule\per\kilo\gram\kelvin}, and 0.82 \unit{\watt\per\meter\kelvin}. These values were taken from~\cite{Cao2019} and assumed to be constant. Finally, the emissivity of glass was 0.90, taken from~\cite{Touloukian1971}.

\newpage

\subsection{Mesh}\label{sec:mesh}

\begin{figure}[h!]
    \centering
    \includegraphics[width=\textwidth]{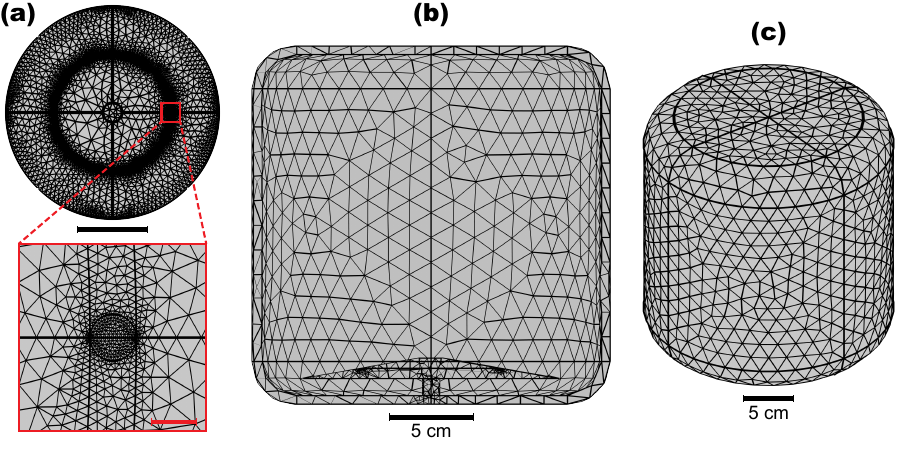}
    \vspace{-0.6cm}
    \caption{Example of a mesh used in our FEM model ($d=1$ \unit{\milli\meter}, $\varphi=10$\unit{\degree}). (a) Top view of the anode, showing the mesh refinement at the edges and around the focal track (scale bar = 5 cm). The zoomed-in image shows the focal spot, where the mesh is the finest (scale bar = 1 mm). (b) Cross-sectional view of the mesh, showing both the anode and the interior of the housing. (c) Exterior of the housing. Details on meshing, including the element sizes in different regions of the X-ray tube, can be found in Sec. \ref{sec:mesh}.}
    \label{fig:mesh}
\end{figure}

In our FEM model, we used a tetrahedral mesh with different element sizes in different regions to account for the steep temperature gradients in and around the focal spot. We performed a mesh refinement study for each focal spot size in the main text ($d=$ 1, 3, and 5 \unit{\milli\meter}) to determine the element sizes that balanced speed and accuracy. In particular, for each $d$-value, we swept over 14 different element sizes spanning an order of magnitude and chose the largest possible element size that resulted in a $\lesssim 1\%$ difference in the focal spot temperature compared with the next largest element size. The results of the mesh refinement study are shown in Table \ref{tab:mesh}, which lists the maximum element sizes in different regions of the X-ray tube. The heading ``Coarse'' refers to the element sizes used to quickly calculate dependence on variables such as $\Omega$, $\varphi$, $U_{\infty}$, and $T_{\text{coolant}}$, which is reported in Sec. \ref{sec:moreresults}. However, even these element sizes led to results that were off from the most refined mesh by not more than a few percents. The minimum element size was $1/10^\text{th}$ the maximum element size in every region except the housing, where it was $1/5^\text{th}$. As an example, Fig. \ref{fig:mesh} shows what the mesh for $d=1$ \unit{\milli\meter} looked like.

\begin{table}[h!]
\begin{threeparttable}
\begin{tabular}{P{1.5in} c c c c}
 \hline\hline
  & \multicolumn{3}{c}{\textbf{Focal spot size [mm]}} &  \\
 \textbf{Region} & 1 & 3 & 5 & \textbf{Coarse}\tnote{*} \\ \hline
 Focal spot surface\tnote{†} & 0.0875 & 0.15 & 0.2 & 0.5 \\
 Focal spot volume\tnote{‡} & 0.35 & 0.6 & 0.8 & 2 \\
 Focal track & 0.7 & 1.2 & 1.6 & 2 \\
 Anode & 16 & 16 & 16 & 16 \\
 Stem & 20 & 20 & 20 & 20 \\
 Housing & 75 & 75 & 75 & 100 \\
 \hline
\end{tabular}
\begin{tablenotes}\footnotesize
\item[*] $d=5$ \unit{\milli\meter}.
\item[†] Boundary mesh.
\item[‡] Region below the surface of the focal spot.
\end{tablenotes}
\end{threeparttable}
\caption{Maximum element size in \unit{\milli\meter} in each region of the X-ray tube for each focal spot size. The minimum element size was $1/10^\text{th}$ the maximum element size in tungsten and $1/5^\text{th}$ in glass.}
\label{tab:mesh}
\end{table}

\clearpage

\section{Additional results}\label{sec:moreresults}

In this section, we provide additional results obtained through quick calculations using a coarse mesh (see Table \ref{tab:mesh}) to better understand the influence of $\Omega$, $\varphi$, $U_{\infty}$, and $T_{\text{coolant}}$. We studied these variables because of their importance to heat transfer and X-ray emission.

\begin{figure}[h!]
    \centering
    \includegraphics[width=\textwidth]{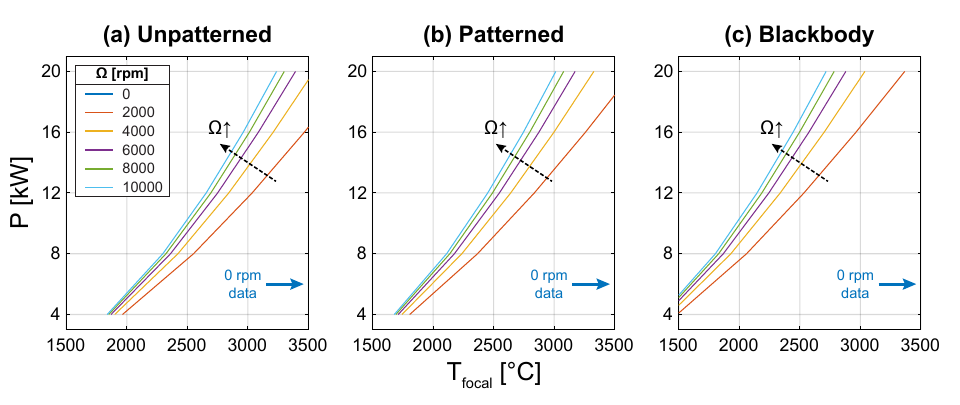}
    \vspace{-0.72cm}
    \caption{Power input $P$ as a function of focal spot temperature $T_{\text{focal}}$ and anode angular velocity $\Omega$. Different colors correspond to different $\Omega$-values in units of rpm (revolutions per minute), and the dashed black arrow indicates the direction of increasing $\Omega$. (a), (b), and (c) show data for the unpatterned, patterned, and blackbody anodes, respectively, i.e., emissivity increases from (a) to (c).}
    \label{fig:omegasweep}
\end{figure}

Figure \ref{fig:omegasweep} shows the power input $P$ as a function of focal spot temperature $T_{\text{focal}}$ and $\Omega$. As expected, $T_{\text{focal}}$ decreases as $\Omega$ increases because $P$ is more evenly distributed around the circumference of the anode. Data for $\Omega=0$ rpm is off the scale, which makes sense since $P$ is in units of \unit{\kilo\watt}---typically, the power rating of a stationary anode X-ray tube is less than 1 \unit{\kilo\watt} for a wide range of clinically relevant exposure times and just over 1--2 \unit{\watt} approaching continuous operation~\cite{Behling2021stationary}.

\begin{figure}[h!]
    \centering
    \includegraphics[width=\textwidth]{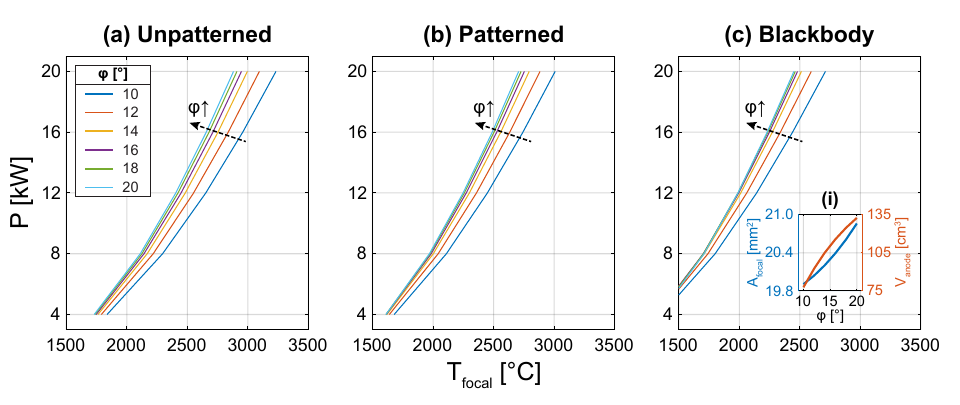}
    \vspace{-0.72cm}
    \caption{$P$ as a function of $T_{\text{focal}}$ and anode angle $\varphi$, as defined in Fig. \ref{fig:sideviewfem}. Different colors correspond to different $\varphi$-values in degrees; the dashed black arrow indicates the direction of increasing $\varphi$. (a)--(c) show data for the unpatterned, patterned, and blackbody anodes, in that order. In (c), the inset (i) shows the focal spot area $A_{\text{focal}}$ (blue) and anode volume $V_{\text{anode}}$ (orange) as a function of $\varphi$ for $d=5$ \unit{\milli\meter}. $A_{\text{focal}}$ is calculated in COMSOL Multiphysics\textsuperscript{\textregistered} to fully account for the curvature of the anode (as opposed to calculating the area of a circle of diameter $d$). $V_{\text{anode}}$ is simply the volume of a truncated cone, of which the base, height, and angle are defined.}
    \label{fig:phisweep}
\end{figure}

Figure \ref{fig:phisweep} shows $P$ as a function of $T_{\text{focal}}$ and $\varphi$. Increasing $\varphi$ increases the focal spot area $A_{\text{focal}}$, which decreases the heat flux and makes the focal spot cooler. This is complemented by an increase in the heat capacity of the anode due to an increase in its volume, $V_{\text{anode}}$. These effects are further illustrated in the inset of Fig. \ref{fig:phisweep}, which explains the reason why the $P$-$T_{\text{focal}}$ curves appear to shift left less and less as $\varphi$ increases. A small change in $\varphi$ leads to a comparatively larger change in $V_{\text{anode}}$ than $A_{\text{focal}}$, and since $V_{\text{anode}}$ as a function of $\varphi$ is concave down, the $P$-$T_{\text{focal}}$ curves ``saturate'' or stop changing very much as a function of $\varphi$.

\begin{figure}[h!]
    \centering
    \includegraphics{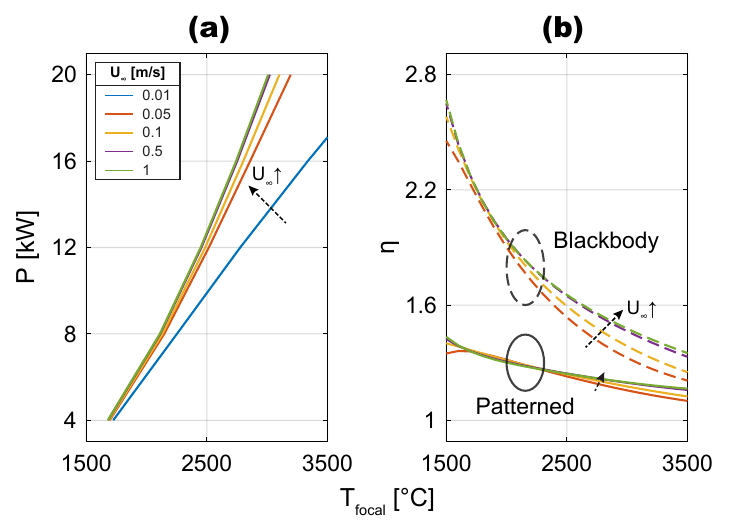}
    \vspace{-0.3cm}
    \caption{$P$ as a function of $T_{\text{focal}}$ and coolant velocity $U_\infty$. (a) As $U_\infty$ decreases, the $P$-$T_{\text{focal}}$ curves shift right in the direction of higher temperatures. This is because the coolant cannot effectively remove all the heat radiated by the anode to the housing. For this reason, the power input enhancement $\eta \equiv P/P_{\text{unpatterned}}$ decreases with $U_\infty$, as shown in (b). The data shown in (a) represent a patterned anode as an example. (b) uses the same legend as (a), and the groups of solid and dashed lines represent patterned and blackbody anodes, respectively. In both (a) and (b), the dashed black arrow indicates the direction of increasing $U_{\infty}$.}
    \label{fig:uinftysweep}
\end{figure}

\begin{figure}[h!]
    \centering
    \includegraphics[width=\textwidth]{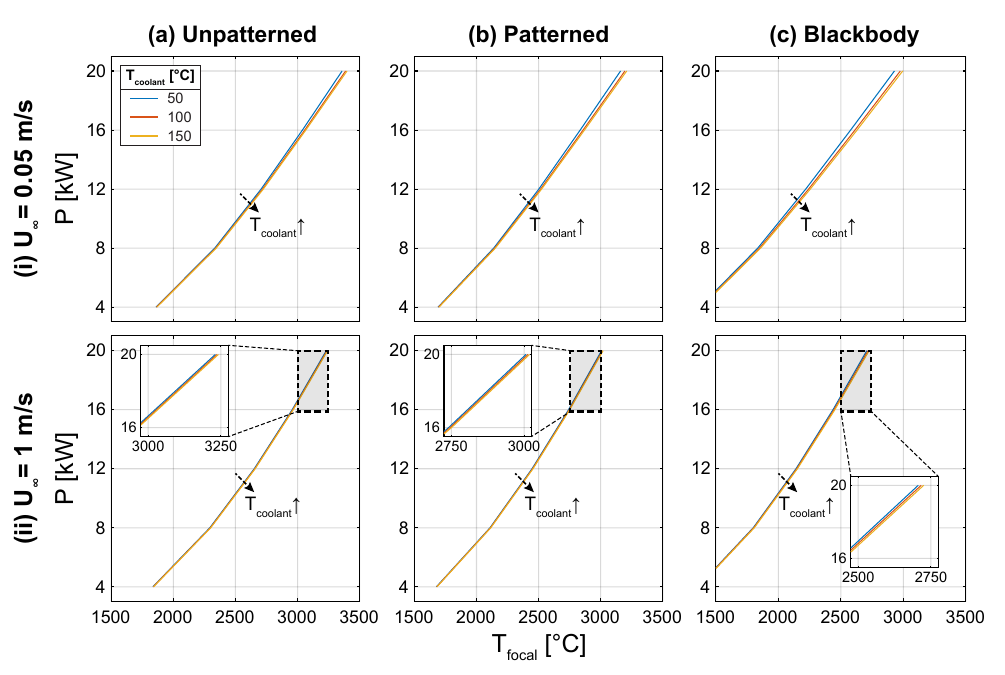}
    \vspace{-0.72cm}
    \caption{$P$ as a function of $T_{\text{focal}}$ and coolant temperature $T_{\text{coolant}}$. Again, (a)--(c) show data for the unpatterned, patterned, and blackbody anodes, and different colors correspond to different values of $T_{\text{coolant}}$. The effect of $T_{\text{coolant}}$ manifests through Eqs. (\ref{eq:housingbc}) and (\ref{eq:finbc}), but it depends on the constants of proportionality $h_{\text{housing}}$ and $h_{\text{fin}}$. In the case of (i) (the first row), the coolant velocity $U_\infty$ is low, so the effect of $T_{\text{coolant}}$, which is to shift the $P$-$T_{\text{focal}}$ curves right as it increases, is noticeable. This effect is caused by the temperature difference between the coolant and the housing---which is proportional to the cooling load---decreasing. In the case of (ii) (the second row), the housing is overcooled, so $h_{\text{housing}}$ is large and $T_{\text{coolant}}$ has little effect.}
    \label{fig:tcsweep}
\end{figure}

Finally, Figs. \ref{fig:uinftysweep}--\ref{fig:tcsweep} show how the $P$-$T_{\text{focal}}$ curves change with $U_\infty$ and $T_{\text{coolant}}$. These variables are extremely important because however much heat is dissipated by thermal radiation (enhanced or not), eventually, it must be carried away by the coolant. In the main text, our calculations assumed the housing was ``overcooled,'' which is the ideal case---all the thermal radiation from the anode and absorbed by the housing is removed. If this assumption is dropped, for example, by decreasing $U_\infty$, nanophotonic thermal management is not as effective because the additional heat emitted by the patterned (or blackbody) anode cannot exit the system via the coolant. In agreement with this intuition, in Fig. \ref{fig:uinftysweep}(a), the power input enhancement $\eta \equiv P/P_{\text{unpatterned}}$ (Eq. (3) in the main text) decreases with $U_\infty$. Figure \ref{fig:uinftysweep} shows how the $P$-$T_{\text{focal}}$ curves look as $U_\infty$ decreases using the patterned anode as an example. As $U_\infty$ increases, an inflection point appears that moves along the $P$-$T_{\text{focal}}$ curves toward lower $P$-values. This makes sense: as the cooling load (from pumping the coolant around the housing) decreases, $P$ begins to saturate as the coolant cannot remove any more heat from the system, resulting in a case where $T_{\text{focal}}$ increases much faster than $P$ and the X-ray tube is likely to fail as a result of overheating. Via Eqs. (\ref{eq:finbc}) and (\ref{eq:housingbc}), increasing $T_{\text{coolant}}$ should have a similar effect to decreasing $U_\infty$ since they work to decrease $q_{\text{fin}}$ and $q_{\text{housing}}$. In Fig. \ref{fig:tcsweep}, changing $T_{\text{coolant}}$ has little effect because $U_\infty$ is large---meaning $h_{\text{fin}}$ and $h_{\text{housing}}$ dominate---and the X-ray tube is overcooled. 
Importantly, there is a practical limit to how hot the coolant can be: if the coolant is transformer oil, it needs to be kept below 200 \unit{\celsius} to prevent carbonization.

\clearpage

\section{Thermal resistance model of the X-ray tube}

\begin{figure}
    \centering
    \includegraphics{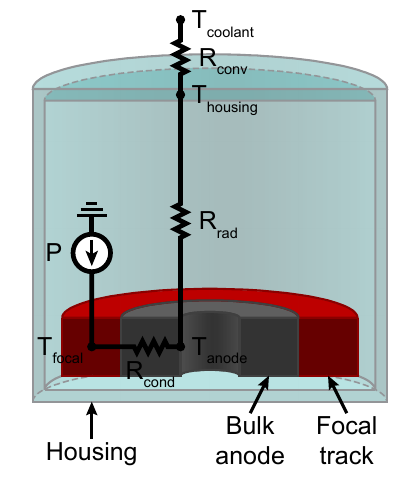}
    \vspace{-0.36cm}
    \caption{Simplified model of the X-ray tube using thermal resistances. Here, the simplified geometry of the X-ray tube is shown along with the thermal circuit diagram and physical locations of resistors and nodes.}
    \label{fig:resistances}
\end{figure}

To understand the dependence of focal spot temperature on key parameters such as power input, focal spot size, and anode angle, we derived an analytical model of the X-ray tube using thermal resistances~\cite{Mills1995,Lienhard2024}. Figure \ref{fig:resistances} shows a simplified version of the X-ray tube. We model the anode as a cylinder and assume it rotates at such a high speed that the temperature of the focal track, $T_{\text{focal}}$, is uniform. The heat entering the focal spot is $P=I_{t}V_{t}$, the product of the tube current and voltage---as previously mentioned, almost all ($>99$\%) of the power input is converted to heat~\cite{Bushong2020,Behling2021}. Heat is conducted from the focal track to the bulk anode, whose temperature is $T_{\text{anode}}$ near the center. $T_{\text{anode}}$, the ``bulk anode temperature,'' can be thought of as the average temperature of the anode excluding the focal track. The conductive thermal resistance from the focal track to the bulk anode is

\begin{equation}\label{eq:condres}
    R_{\text{cond}}=\frac{\ln\left(\frac{R-d\sec\varphi}{r}\right)}{2\pi k_{\text{W}}t},
\end{equation}

\noindent where $R$ is the radius of the anode, $d$ is the focal track size, $\varphi$ is the anode angle, $r$ is the radius of the stem, and $t$ is the thickness of the anode. In addition, $k_{\text{W}}$ is the thermal conductivity of tungsten which can be temperature-dependent (as per Refs. ~\cite{Incropera2007,Šmid1993} and Fig. \ref{fig:wthermophys}).

Next, if the surface area of the focal track is much less than that of the bulk anode, we can assume (in fact, show using radiative resistances~\cite{Mills1995,Lienhard2024}) that radiation from the focal track is negligible compared to radiation from the anode. Therefore, the radiative thermal resistance $R_{\text{rad}}$ is connected in series with $R_{\text{cond}}$. This is given by:

\begin{equation}\label{eq:radres}
    R_{\text{rad}}=\frac{1}{h_{\text{rad}}A_{\text{anode}}}.
\end{equation}

\noindent Here, $A_{\text{anode}}=\pi(R-d)^2 - \pi r^2$ and $h_{\text{rad}}$ is the temperature-dependent radiative heat transfer coefficient:

\begin{equation}\label{eq:radhtc}
    h_{\text{rad}}=4\epsilon\sigma\left(\frac{T_{\text{anode}} + T_{\text{housing}}}{2}\right)^3,
\end{equation}

\noindent where $\epsilon$ is the total hemispherical emissivity of the anode, $\sigma$ is the Stefan-Boltzmann constant, and $T_{\text{housing}}$ is the temperature of the housing, which receives the radiation emitted by the bulk anode. Technically, Eq. (\ref{eq:radhtc}) is a useful approximation that makes the resulting system of equations amenable to an iterative solution. Details on $h_{\text{rad}}$ can be found in~\cite{Glicksman}; it is straightforward to derive.

Finally, the housing is liquid cooled. Assuming the housing is thin (such that its temperature changes little from the inside to the outside), the convective thermal resistance is

\begin{equation}\label{eq:convres}
    R_{\text{conv}}=\frac{1}{h_{\text{conv}}A_{\text{housing}}},
\end{equation}

\noindent where $h_{\text{conv}}$ is the convective heat transfer coefficient and $A_{\text{housing}}$ is the surface area of the housing. If the diameter of the housing is $D$ and the height is $H$, $A_{\text{housing}}=\frac{\pi D^{2}}{4}H$. $h_{\text{conv}}$ can be estimated using Eq. (\ref{eq:nustem}), for example, one of the heat transfer correlations used in our finite element method model.



The thermal circuit diagram is shown in Fig. \ref{fig:resistances}(b). The thermal resistances given by Eqs. (\ref{eq:condres}), (\ref{eq:radres}), and (\ref{eq:convres}) are connected in series. If the temperature of the coolant is $T_{\text{coolant}}$, the relationship between $T_{\text{focal}}$ and $P$ is

\begin{equation}\label{eq:Tfocalmodel}
    P = \frac{T_{\text{focal}} - T_{\infty}}{\frac{\ln\left(\frac{R}{R-d}\right)}{2\pi k_{\text{W}}t} + \frac{1}{4\epsilon\sigma\left(\frac{T_{\text{anode}} + T_{\text{housing}}}{2}\right)^{3}\left[\pi(R-d)^2 - \pi r^2\right]} + \frac{1}{h_{\text{conv}}\frac{\pi D^{2}}{4}H}}.
\end{equation}

\noindent In other words, $P$ equals the temperature difference divided by the sum of thermal resistances in series between the nodes, analogous to electrical circuits. Equation (\ref{eq:Tfocalmodel}) is useful for deriving scaling laws since it depends on almost all the design variables of the X-ray tube relevant to heat transfer (to be discussed). In practice, the temperatures are computed using simpler equations.

\subsection{Solution}

Mathematically, the thermal resistance model comprises a system of nonlinear equations in $T_{\text{focal}}$, $T_{\text{anode}}$, and $T_{\text{housing}}$. The temperature dependence of the material properties, along with the inclusion of radiation which depends on temperature to the fourth power, make the system of equations nonlinear. We used the same material properties as our FEM model (see Sec. \ref{sec:materials}, in particular Figs. \ref{fig:wthermophys}--\ref{fig:oilthermophys}). $k_{\text{W}}$ was evaluated at $\overline{T} = (T_{\text{focal}} + T_{\text{anode}})/2$ to reflect the average temperature of the anode from the focal track to the center. $\epsilon$ was evaluated at $T_{\text{anode}}$. $k_{\text{oil}}$, $\nu_{\text{oil}}$, and $\alpha_{\text{oil}}$ were evaluated at the film temperature, $T_{\text{film}}=\left(T_{\text{housing}} + T_{\infty}\right)/2$.

We solved for $T_{\text{focal}}$, $T_{\text{anode}}$, and $T_{\text{housing}}$ using an iterative method coded in MATLAB\textsuperscript{\textregistered}. Since the thermal resistances are in series, starting from an initial guess, we can compute the temperatures in sequence starting from $T_{\text{housing}}$:

\begin{equation}\label{eq:Thousingsolution}
    T_{\text{housing}}=T_{\text{coolant}} + PR_{\text{conv}}.
\end{equation}

\noindent The material properties are computed using $T_{\text{housing}}$ from the previous iteration  (because they depend on $T_{\text{film}}$), which means everything on the right-hand side of Eq. (\ref{eq:Thousingsolution}) is known. Similarly, $T_{\text{anode}}$ is given by

\begin{equation}\label{eq:Tanodesolution}
    T_{\text{anode}}=T_{\text{housing}} + PR_{\text{rad}}.
\end{equation}

\noindent The right-hand side of Eq. (\ref{eq:Tanodesolution}), which depends on $T_{\text{anode}}$ via $R_{\text{rad}}$ (see Eqs. (\ref{eq:radres}) and (\ref{eq:radhtc})), can be computed using $T_{\text{anode}}$ from the previous iteration. Alternatively, $R_{\text{rad}}$ (or rather $h_{\text{rad}}$) can be expanded in terms of $T_{\text{anode}}$, and Eq. (\ref{eq:Tanodesolution}) can be directly solved as a polynomial. This is easily done using built-in functions in MATLAB\textsuperscript{\textregistered}, and we confirmed that both approaches (linear approximation versus polynomial) produce almost identical results. Finally, $T_{\text{focal}}$ can be computed using the equation

\begin{equation}\label{eq:Tfocalsolution}
    T_{\text{focal}}=T_{\text{anode}} + PR_{\text{cond}}.
\end{equation}

\noindent Equations (\ref{eq:Thousingsolution})--(\ref{eq:Tfocalsolution}) are evaluated until the current iteration changes by less than 1\% compared to the previous iteration.

\subsection{Scaling laws}

Since $P$ is limited by $T_{\text{focal}}$, the hottest temperature on the surface of the anode, it is useful to simplify and inspect Eq. (\ref{eq:Tfocalmodel}) to understand how $T_{\text{focal}}$ scales with key parameters. The first step is to approximate the transcendental functions in Eq. (\ref{eq:Tfocalmodel}) by Taylor polynomials. Since the term $(R-d\sec\varphi)/r$ is on the order of $10^0$, we can approximate $R_{\text{cond}}$ by

\begin{equation}\label{eq:condresapprox}
    R_{\text{cond}} \approx \frac{\frac{R-d\left(1+\frac{\varphi^2}{2}\right)}{r}-1}{2\pi k_{\text{W}}t},
\end{equation}

\noindent using the Taylor polynomials $\ln x = (x-1) + O\left[(x-1)^{2}\right]$ and $\sec\varphi = 1 + \varphi^{2}/2 + O\left(\varphi^{3}\right)$. Substituting Eq. (\ref{eq:condresapprox}) into Eq. (\ref{eq:Tfocalmodel}), we obtain the approximation

\begin{multline}\label{eq:Tfocalapprox}
    T_{\text{focal}} \approx T_{\infty} \\
    + P\left\{\frac{\frac{R-d\left(1+\frac{\varphi^2}{2}\right)}{r}-1}{2\pi k_{\text{W}}t} + \frac{1}{4\epsilon\sigma\left(\frac{T_{\text{anode}} + T_{\text{housing}}}{2}\right)^{3}\left[\pi(R-d)^2 - \pi r^2\right]} + \frac{1}{h_{\text{conv}}\frac{\pi D^{2}}{4}H}\right\}.
\end{multline}

\newpage

By inspecting Eq. (\ref{eq:Tfocalapprox}), we identify a number of useful scaling laws:

\begin{itemize}
    \item $T_{\text{focal}} \sim P$. This makes sense: as the power input increases, the focal track temperature increases. In practice, we expect the dependence to be nonlinear due to (1) radiation depending on temperature to the power of four and (2) temperature-dependent material properties. This is what we observed in both our thermal resistance model and our finite element method model.
    \item $T_{\text{focal}} \sim \epsilon^{-1}$. As the total hemispherical emissivity of the anode increases, radiation increases, which cools the anode and decreases the focal track temperature. This relationship between $T_{\text{focal}}$ and $\epsilon$ justifies the use of nanophotonics to improve thermal management.
    \item $T_{\text{focal}} \sim R-d$ and $T_{\text{focal}} \sim 1/\left[(R-d)^{2} - r^{2}\right]$. In this model, conduction and radiation have competing effects as a function of $d$ because (1) the sizes of the focal track and bulk anode are coupled (see Fig. \ref{fig:resistances}) and (2) the length scales associated with conduction, radiation, and heat flux entering the focal spot are different. This is consistent with our observations in the main text.
    \item $T_{\text{focal}} \sim \varphi^2$. This is one aspect of our thermal resistance model that is not consistent with our finite element method model. According to Fig. \ref{fig:phisweep}, increasing $\varphi$ should drive $T_{\text{focal}}$ down. However, this is because the thermal resistance model does not account for the heat capacity of the anode.
\end{itemize}

\begin{figure}
    \centering
    \includegraphics{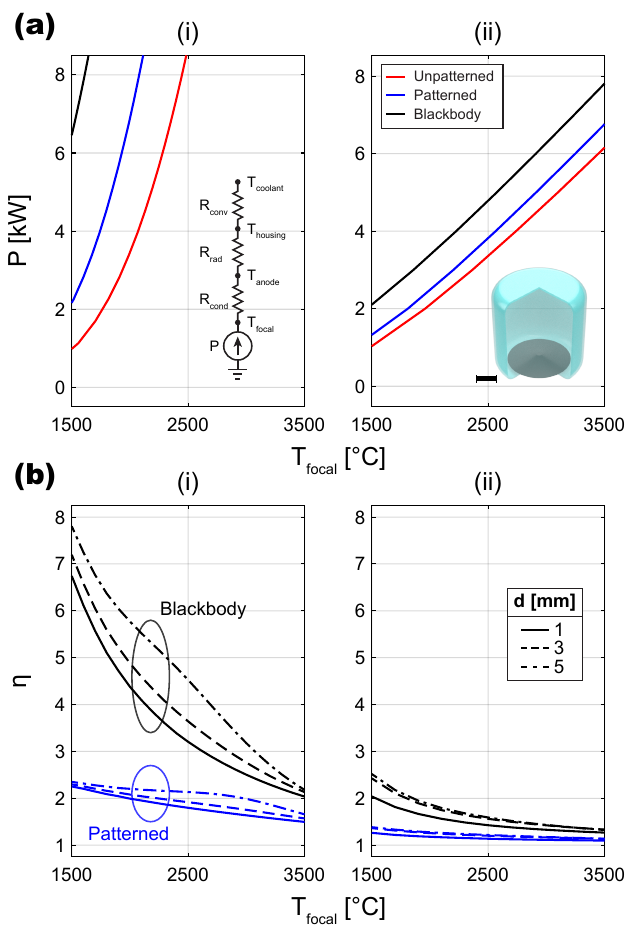}
    \vspace{-0.45cm}
    \caption{The thermal resistance model shows qualitative good (even semi-quantitative) agreement with the general trends observed in the FEM model. (a) Power input $P$ as a function of $T_{\text{focal}}$. (i) Calculated using the thermal resistance model, showing good agreement with the finite element method model in the main text, reproduced in (ii). Inset (i): ``unfolded'' thermal circuit diagram, reproduced from (a). Inset (ii): render of the model geometry (scale bar = 5 \unit{\centi\meter}). (b) Power input enhancement $\eta$ calculated using (i) the thermal resistance model and (ii) the FEM model. The thermal resistance model predicts higher $\eta$-values but agrees with the FEM model in terms of how $\eta$ depends on $T_{\text{focal}}$ and $d$.}
    \label{fig:analvsnum}
\end{figure}

\subsection{Comparison to finite element method model}

Our thermal resistance model agrees semi-quantitatively with our finite element model, meaning (1) it captures the general trends we observed in the main text and Sec. \ref{sec:moreresults} and (2) the order of magnitude of is accurate. In Fig. \ref{fig:analvsnum}(a), we plot $P$ as a function of $T_{\text{focal}}$, calculated using (i) thermal resistances and (ii) the finite element method (reproduced from Fig. 2(a) in the main text). The former shows good agreement with the latter, in that $P$ and $T_{\text{focal}}$ have a positive, concave up relationship. Figure \ref{fig:analvsnum}(b) shows the power input enhancement $\eta$ as a function of $T_{\text{focal}}$. Here, our thermal resistance model still captures the general trends as a function of emissivity (i.e., patterned versus blackbody) and $d$, but it predicts $\eta$-values that are several times higher than our finite element method model. We believe this is because our thermal resistance model does not (and cannot) account for the heat capacity of the anode. Some heat is stored in the anode rather than being dissipated. This helps with thermal management by providing another avenue for the heat generated by the electron beam, decreasing $T_{\text{focal}}$ and cooling down the focal spot but preventing heat from exiting the system via thermal radiation, thus decreasing $\eta$.

\clearpage

\section{Geant4 model of bremsstrahlung}

\begin{figure}[h!]
  \includegraphics[width=4.8in]{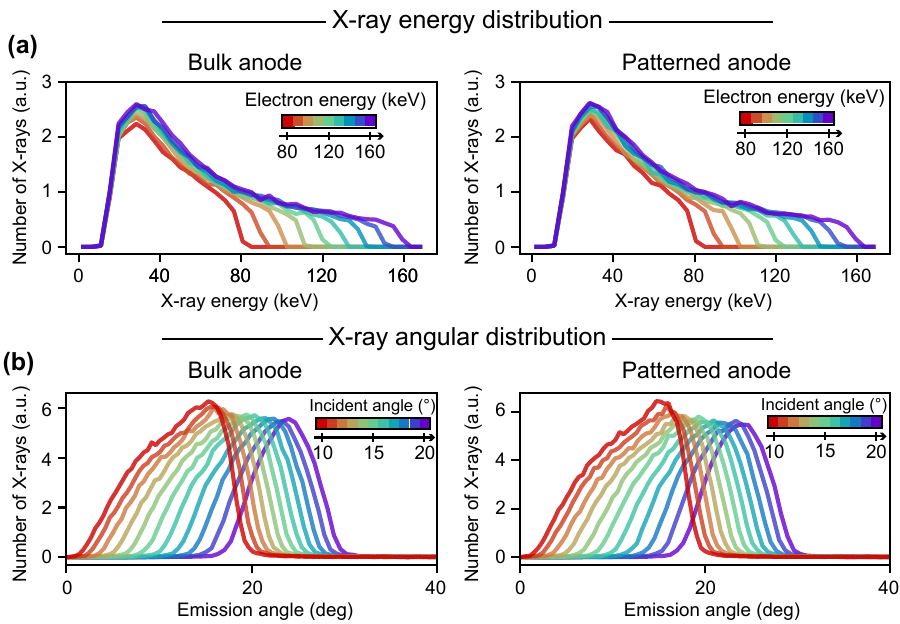}
  \vspace{-0.3cm}
  \caption{Energy and angular spectrum of X-rays generated from different types of anodes. (a) Energy spectrum of X-ray photons generated under different incident electron energies for the bulk anode (left panel) and the patterned anode (right panel). X-ray photons are collected by recording their energies with spherical detectors in the Geant4 simulation. Then, low energy X-ray photons are filtered out assuming that a 0.1~\unit{\centi\meter} thick aluminum plate is attached in front of the detector, which is a typical filtering approach in X-ray tubes. (b) Angular spectrum of X-ray photons under different incident angles of electron beams for the bulk anode (left panel) and patterned anode (right panel). }
  \label{fig:geant4}
\end{figure}

We use Monte Carlo simulations in Geant4~\cite{Agostinelli2003geant4} to confirm a minimal change in bremsstrahlung emission as a result of patterning. In our Geant4 simulations, we model a simple X-ray tube by bombarding a tungsten anode with free electrons. Part of the energy of the free electrons is converted to X-ray photons via bremsstrahlung. Then, the emitted X-ray photons are detected using a spherical detector surrounding the tungsten anode, recording both the position and the energy of the X-ray photons. To compare bremsstrahlung emitted by different tungsten anodes, we prepare a bulk anode and a patterned anode with equal sizes (0.1\unit{\milli\meter} $\times$ 0.1\unit{\milli\meter} $\times$ 0.1\unit{\milli\meter}). We then calculate the energy and angular distribution of X-rays generated from these two different anodes. Figure~\ref{fig:geant4} demonstrates that replacing the bulk anode with a patterned anode introduces a minimal change in both the energy spectrum and the angular spectrum of X-rays. The patterned anode cannot enhance the electron-to-X-ray conversion efficiency because X-ray generation via bremsstrahlung relies on the stopping power of the anode, which is an inherent property of the anode material. Therefore, we can conclude that the ability yield more X-rays from the patterned anode results from improved thermal management with nanophotonic structures, not from an enhancement of bremsstrahlung. Furthermore, this shows that the pattern does not significantly affect the interaction of electrons with the anode.

